\documentclass{aa}  

\usepackage{graphicx}
\usepackage[varg]{txfonts}
\usepackage[normalem]{ulem}
\usepackage[usenames]{color}
\usepackage{mathrsfs}
\usepackage{natbib}
\usepackage{enumitem}
\usepackage{hyperref}
\usepackage{geometry}                           
\usepackage{graphicx}                           
\usepackage{amssymb}
\usepackage{nccmath}
\usepackage{booktabs}
\usepackage[tableposition=top]{caption} 
\usepackage{multirow}
\usepackage{hyperref}
\usepackage[font=small,labelfont=bf]{caption}
\usepackage{media9}

\usepackage{txfonts}
\usepackage{xcolor}
\usepackage{soul}

\newcommand{\Mm}{{\mathrm{\, Mm}}}
\newcommand{\kms}{{\mathrm{\, km \,s^{-1}}}}

\newcommand{\secs}{{\mathrm{\, seconds}}}
\newcommand{\mins}{{\mathrm{\, minutes}}}

\definecolor{valeriia}{rgb}{1.0, 0.0, 0.0}
\definecolor{manuelcomment}{rgb}{0.93, 0.57, 0.13}
\definecolor{manuel}{rgb}{0.3, 0.3, 0.9}
\definecolor{elena}{rgb}{0.0,0.5,0.0}

\DeclareMathOperator\erf{erf}

\begin{document} 

\title{Numerical simulations of prominence oscillations triggered by external perturbations}

   \author{V. Liakh\inst{1}
   \and 
   M. Luna \inst{2, 3}
   \and 
  E. Khomenko\inst{4, 5}}

\institute{Centre for mathematical Plasma Astrophysics, Department of Mathematics,\\
KU Leuven, Celestijnenlaan 200B, 3001 Leuven, Belgium
              \email{valeriia.liakh@kuleuven.be}
         \and Departament de F\'{\i}sica, Universitat de les Illes Balears, E-07122, Palma de Mallorca, Spain
         \and Institute of Applied Computing \& Community Code (IAC$^3$), UIB, Spain
         \and Instituto de Astrof\'{\i}sica de Canarias, E-38205 La Laguna, Tenerife, Spain
         \and Departamento de Astrof\'{\i}sica, Universidad de La Laguna, E-38206 La Laguna, Tenerife, Spain
          }
   

 \date{}

 \titlerunning{Numerical simulations of LAOs}
\authorrunning{Liakh, Luna \& Khomenko}

\abstract
   {Several energetic disturbances have been identified as triggers of the large-amplitude oscillations (LAOs) in prominences. Observations show that Moreton or EIT waves excite prominence oscillations of the longitudinal, transverse, or mixed polarities. However, the mechanisms for the LAOs excitation by these waves are not well understood.}
   {In this study, we aim to investigate mechanisms for LAOs triggering by self-consistent perturbation produced by an eruption and by energetic waves caused by a distant energy source.}
   {We perform time-dependent numerical simulations in 2.5D and 2D setups using magnetohydrodynamic (MHD) code MANCHA3D, involving a flux rope and dipped arcade magnetic configurations with an artificially loaded prominence mass in the magnetic dips. Two types of disturbances are applied to excite prominence oscillations. The first type involves perturbations produced self-consistently by an eruption, while the second type of perturbation is associated with the waves caused by an artificial energy release.}
   {In the simulation with the eruption, we obtain that this eruption by itself does not produce LAOs in the prominence located in its vicinity. Its only effect is in inclining the magnetic configuration of the prominence. While the erupting flux rope rises, an elongated current sheet forms behind it. This current sheet becomes unstable and breaks into plasmoids. The downward-moving plasmoids cause perturbations in the velocity field by merging with the post-reconnection loops. This velocity perturbation propagates in the surroundings and enters the flux rope, causing the disturbance of the prominence mass. The analysis of the oscillatory motions of the prominence plasma reveals the excitation of small-amplitude oscillations (SAOs), which are a mixture of longitudinal and vertical oscillations with short and long periods. 
   	In the simulation with a distant artificial perturbation, a fast-mode shock wave is produced, and it gradually reaches two flux rope prominences at different distances. This shock wave excites vertical LAOs, and longitudinal SAOs with similar amplitudes, periods, and damping times in both prominences. 
   	Finally, in the experiment with the external triggering of LAOs of solar prominences by an artificial perturbation in a dipped arcade prominence model, we find that, although the vector normal to the front of a fast-mode shock wave is parallel to the spine of the dipped arcade well before the contact, this wave does not excite longitudinal LAOs. When the wave front approaches the prominence, it pushes the dense plasma down, establishing vertical LAOs, and motions due to compression and rarefaction along the magnetic field.}
   {The external triggering of prominence oscillations is a complex process that excites LAOs or SAOs of the longitudinal or transverse polarizations or a mix of both types. It is not an easy task to produce LAOs in prominences because the triggering event should have a sufficient amount of energy. The orientation of the prominence axis with respect to the driving event may play a crucial role in triggering a certain type of LAOs.}
   \maketitle
%
 %

\section{Introduction}\label{sec:introduction}

Solar prominences are cool and dense plasma clouds located often inside the magnetic dips high in the solar corona and supported against gravity by the magnetic force. They are highly dynamic structures. One of the manifestations of their dynamics is prominence oscillations, classified as small-amplitude (SAOs) and large-amplitude oscillations (LAOs) according to their velocity with a threshold $10\kms$ \citep[see reviews by][]{Oliver:2002solphys,Arregui:2018spr}. Prominence oscillations can also be classified according to the direction of plasma motions with respect to the magnetic field. Longitudinal prominence oscillations mean periodic motion along the magnetic field. On the contrary,  oscillations with velocity directed perpendicular to the magnetic field are called transverse oscillations. More information on the classification and properties of the different types of oscillations can be found in the last update of the living review by \citet{Arregui:2018spr}. 

LAOs should have a huge amount of energy due to the combination of a large mass and large velocities involved. It is intriguing how perturbations can transport, and prominence can accumulate such an amount of energy.
Most of the LAOs events have been associated with energetic disturbances. One piece of the earliest evidence of such kind of a disturbance was by \citet{Moreton:1960pasp}, who discovered the existence of waves emanating from flares or coronal mass ejections (CME) and propagating in the solar atmosphere with huge velocities of $500-1500\kms$. More recently, the Extreme ultraviolet Imaging Telescope (EIT) aboard the Solar and Heliospheric Observatory (SOHO) discovered another wave-like phenomenon in the solar corona, now called EIT wave, according to the name of the instrument, or EUV wave. \citep{Moses:1997solphys,Thompson:1998grl}. These EIT waves propagate at speeds of $200-300\kms$ and are usually associated with CMEs or eruptions \citep{Biesecker:2002apj,Chen:2006apj,Chen:2011spr}. The Moreton wave is widely accepted as a fast-mode wave that propagates in the chromosphere and the low corona. Initially, EIT waves were explained as a coronal counterpart of the corresponding Moreton waves \citep{Thompson:1998grl,Wang:2000apj}. However, this interpretation contradicted the observed velocities of these waves. EIT waves usually have approximately three times smaller velocities than the Moreton waves. More importantly, it has been found in observations that the EIT fronts can be stationary, implying that the fronts can be trapped in the magnetic arcades slowing down there, and finally stopping close to quasi-separatrix layers \citep[QSL; ][]{Delannee:1999solphys,DeVore:2000apj}. Since the fast-mode is expected to propagate through separatrix, this behavior suggests that EIT waves have a different physical nature. Different explanations have been proposed for the EIT waves, such as the magnetic field line stretching model \citep{Chen:2002apjl,Chen:2005apj}, successive reconnection model \citep{Attrill:2007asn}, the slow-mode wave model \citep{Wills-Davey:2007apj}, and the current shell model \citep{Delannee:2008solphys}. \cite{Chen:2002apjl, Chen:2005apj} performed a numerical study of an erupting flux rope and discussed a possible origin of the Moreton and EIT waves. Their numerical experiments showed a piston-like shock straddling the flux rope, whose skirt sweeps the solar surface and propagates with a velocity of more than $700\kms$. It was suggested that the skirt of the shock wave corresponds to the coronal Moreton wave. Another wave-like phenomenon, propagating with a smaller velocity of around $200\kms$, was found in these simulations. The authors associated this phenomenon with an EIT wave formed by successive stretching and opening of the field lines covering the erupting flux rope. The region of the enhanced density was observed in front of the stretching field lines, forming this wave, while its inner region behind the EIT wave coincides with dimmings. Observations confirmed that the EUV wave has two components associated with a fast-mode wave and a non-wave phenomenon \citep{Chen:2011apj}. Following a recent classification by \citet{Chen:2016solphys}, the fast-mode EUV wave is the coronal Moreton wave, whereas the slow non-wave phenomenon is the so-called EIT ``wave''. 

The evidence for the LAOs excitation in filaments by Moreton and EIT waves has been shown in many observations \citep{Eto:2002pasj, Okamoto:2004apj, Gilbert:2008apj, Asai:2012apjl, RiuLiu:2013apj, Shen:2014apj1, Xue:2014solpol, Takahashi:2015apj, Shen:2017apj}. Simultaneous excitation of the different LAO polarizations by the same waves from energetic events has been observed. \citet{Gilbert:2008apj} described an event in which the filament oscillations excited by the Moreton wave had a mixed behavior, showing different polarizations.  
More recent observations of the LAOs excitation in filaments by Moreton and EIT waves have been reported by \citet{Pant:2016solphys}, \citet{ Wang:2016apjl}, \citet{Zhang:2017apj}, and \citet{Mazumder:2020aap}. 

An interesting example of LAOs excitation was reported by \citet{Shen:2014apj2}. A shock wave, associated with a powerful flare, excited the transverse oscillations in the filament and the prominence and the longitudinal oscillations in another filament. The authors proposed that, depending on the direction between the arrival of the wave and the filament spine, the shock wave could excite longitudinal or transverse oscillations. If the vector normal to the wave front is transverse with respect to the filament spine at the moment of the wave arrival, oscillations of horizontal or vertical polarization can be excited. Longitudinal oscillations can be triggered when the vector normal to the front points in the direction parallel to the filament spine. However, as shown by observations by \citet{Shen:2014apj1}, this explanation seems insufficient since some filaments remain unperturbed, despite being located at a close distance from a flare, while the same perturbation triggered a chain of solar filaments.
 While the energetic coronal waves excite predominantly vertical oscillation, the excitation of longitudinal oscillations is often related to the magnetic activity in the filament vicinity, such as magnetic reconnection resulting in jets \citep{Luna:2014apj,Zhang:2017apj}, flare activity \citep{Jing:2003apjl,Jing:2006solphys,Vrsnak:2007aap,Li:2012apj,Zhang:2020aap1}, filament eruptions \citep{Isobe:2006aap,Isobe:2007solphys}.

From the theoretical point of view, propagation and interaction of Moreton and EIT waves with the surrounding magnetic arcades, coronal holes, and current-like prominences were studied in many works \citep[see, e.g.,][]{Chen:2002apjl,Chen:2005apj,Vrsnak:2016solphys,Chen:2016solphys,Piantschitsch:2017apj,Afanasyev:2018aap,Mei:2020mnras,Zurbriggen:2021solphys}.
\citet{Chen:2016solphys} studied the interaction between a fast-mode wave and an isolated magnetic flux system by means of numerical simulations. The authors showed that when the wave impacts the magnetic structure and reaches the region of plasma-$\beta$ close to unity, part of the fast-mode wave is converted to the slow-mode wave. This slow wave stays trapped inside the magnetic loops, forming a non-propagating front detected in observations and called stationary EUV front \citep{Chandra:2016apj}. Although these numerical simulations gave essential information about the waves associated with the energetic events, many aspects of the interaction of these waves with the close and distant prominence mass and excitation of LAOs remain unclear.

In this paper, we study the possibility of triggering LAOs by external perturbations in three different numerical experiments. 
In numerical experiments, we distinguish two ways of producing the energy perturbations: self-consistently and artificially. We call self-consistent simulations when the energetic event that produces the perturbation is also simulated in the experiment. On the other hand, we call artificial perturbations when it is imposed in the experiment.
Recently, \citet{luna_large-amplitude_2021} conducted a numerical experiment of an energy release process in a magnetic structure, resulting in excitation of the large-amplitude longitudinal oscillations (LALOs). This is an example of a self-consistent experiment. In contrast, \citet{Liakh:2020aap} employed the triggering by an artificial perturbation in order to mimic a real energetic disturbance, such as a distant flare, placed at some region of the numerical domain.
Here we first study the scenario where a nearby flux rope eruption causes external perturbation. After that, we show the experiment where an energetic wave is produced artificially and impacts two flux rope prominences or one prominence hosted by the dipped arcade magnetic field.

This paper is organized as follows: in Sect. \ref{sec:eruption}, we describe the experiments where the external perturbation is self-consistent and caused by an eruption of a nearby flux rope.
After that, in Sect. \ref{sec:two-prominences}, we show the experiment where an energetic wave is produced artificially and impacts two flux rope prominences placed at different distances from the perturbation location. 
In Sect. \ref{sec:dipped-arcade}, we describe the interaction of a distant artificial disturbance with a dipped arcade magnetic field and the prominence located in the magnetic dips. In Sect. \ref{sec:summary}, we summarize the main findings.

\section{Self-consistent perturbation associated with an erupting flux rope}\label{sec:eruption}

In this experiment, we study the triggering of the prominence oscillations by the neighboring eruption by means of 2.5D numerical simulations using the magnetohydrodynamic (MHD) code MANCHA3D \citep{Khomenko:2006apj,Felipe:2010apj,Khomenko:2012apj}. 
The eruption in this numerical setup is based on the standard solar flare model, whose main ingredients are the flux rope and magnetic reconnection underneath it \citep[see, e.g.][]{Martens:1989solphys}. 

\subsection{Numerical model}

\begin{figure*}[!ht]
	\centering\includegraphics[width=0.85\textwidth]{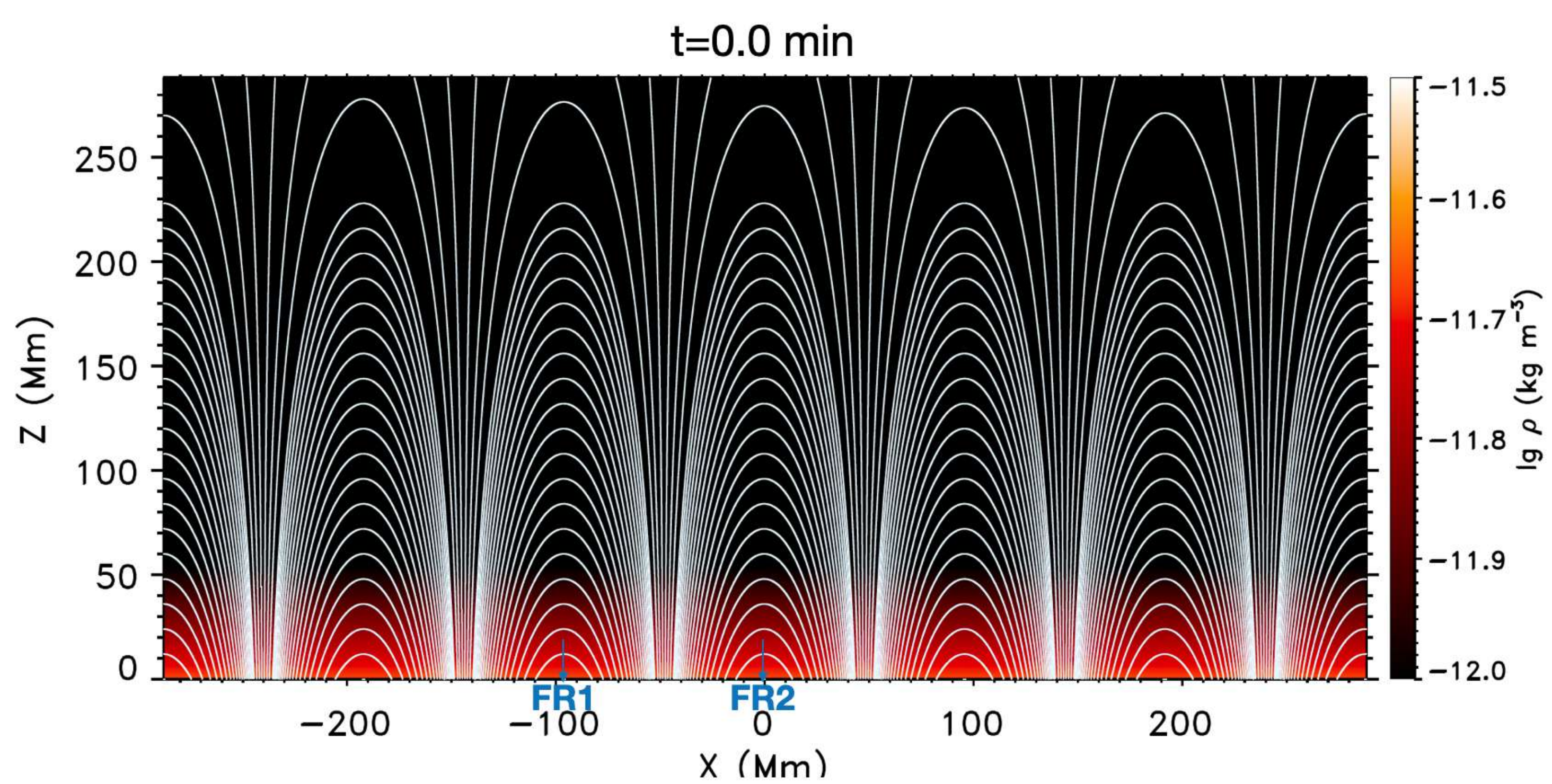}
	\caption{Density distribution and magnetic field configuration in the initial atmosphere in the lower part of the numerical domain. The blue arrows mark the positions where the flux ropes will be formed after the boundary driving process. \label{fig:eruption-initial}}
\end{figure*}

The numerical domain consists of $2400\times 3200$ grid points, corresponding to the physical size of $576\Mm\times 768 \Mm$ (a spatial resolution of $0.24\Mm$). We use a Cartesian coordinate system with $z$- and $x$-axes corresponding to vertical and horizontal directions, respectively. 
Part of the numerical domain is shown in Fig. \ref{fig:eruption-initial}, where the vertical extension is shown up to $z=290\Mm$ instead of the $768\Mm$ of the entire domain. The flux rope hosting the prominence (FR1) will be formed at $x=-96\Mm$, and the erupting flux rope (FR2) will be formed at $x=0\Mm$. 

The initial atmosphere is an isothermal corona with a temperature of $T=1.5\ $ MK, gravitationally stratified along the vertical direction with a density at the bottom of the numerical domain of $\rho=2.12\times 10^{-12}\ \mathrm{kg\ m^{-3} }$. The non-adiabatic effects, such as thermal conduction and optically thin radiation, have not been incorporated into the simulations. However, we include Newtonian cooling during the FR1 formation stage to avoid excessive heating inside it and keep plasma-$\beta$ lower than unity. A more detailed description of the use of Newtonian cooling during the formation of the prominence hosting flux rope can be found in \citet{Liakh:2020aap}. The corresponding cooling term in the energy equation is defined later. 

The initial atmosphere is permeated by a force-free magnetic field represented by a periodic sheared arcade.
The magnetic field components are defined as follows \footnote{Correcting on typos in Eqs. 1-3 of \citet{Liakh:2020aap}},
\begin{eqnarray}\label{magnetic_components}
	B_{x}&=&-\frac{2LB_{0}}{\pi H_{b}}\cos{\frac{\pi x}{2L}} \exp\left({-\frac{z}{H_{b}}}\right) \, , \\ \label{magnetic_component_y}
	B_{y}&=&-\sqrt{1-\left(\frac{2L}{\pi H_{b}}\right)^2}B_{0}\cos{\frac{\pi x}{2L}}\exp\left({-\frac{z}{H_{b}}}\right) \, ,\\ \label{magnetic_component_z}
	B_{z}&=&B_{0}\sin{\frac{\pi x}{2L}}\exp\left({-\frac{z}{H_{b}}}\right) \, ,
\end{eqnarray}
where $L=48\Mm$ is half of the lateral extension of each arcade, $H_b=118\Mm$ is the magnetic scale height, and $B_0=10$ G is the magnetic field strength at the bottom. The shear angle is defined as the angle formed by the field vector with the $xz$-plane as
\begin{equation}\label{eq:methodology-tantheta}
	\tan{\theta}=\frac{B_y}{B_x}=\sqrt{\left(\frac{\pi H_{b}}{2L}\right)^2-1}  \, .
\end{equation}
Using the previous parameters, we obtain the shear angle $\theta=75\,^{\circ}$, indicating that the field is quite aligned with the $y$-axis.
The initial magnetic field is shown in Fig. \ref{fig:eruption-initial} in a 2D projection. We note that several identical magnetic arcades are nested in the numerical domain. As we use the periodic system, we increase the horizontal extension of the domain in order to reduce the perturbation coming from the side boundaries.

\begin{figure*}[!ht]
	\centering\includegraphics[width=0.85\textwidth]{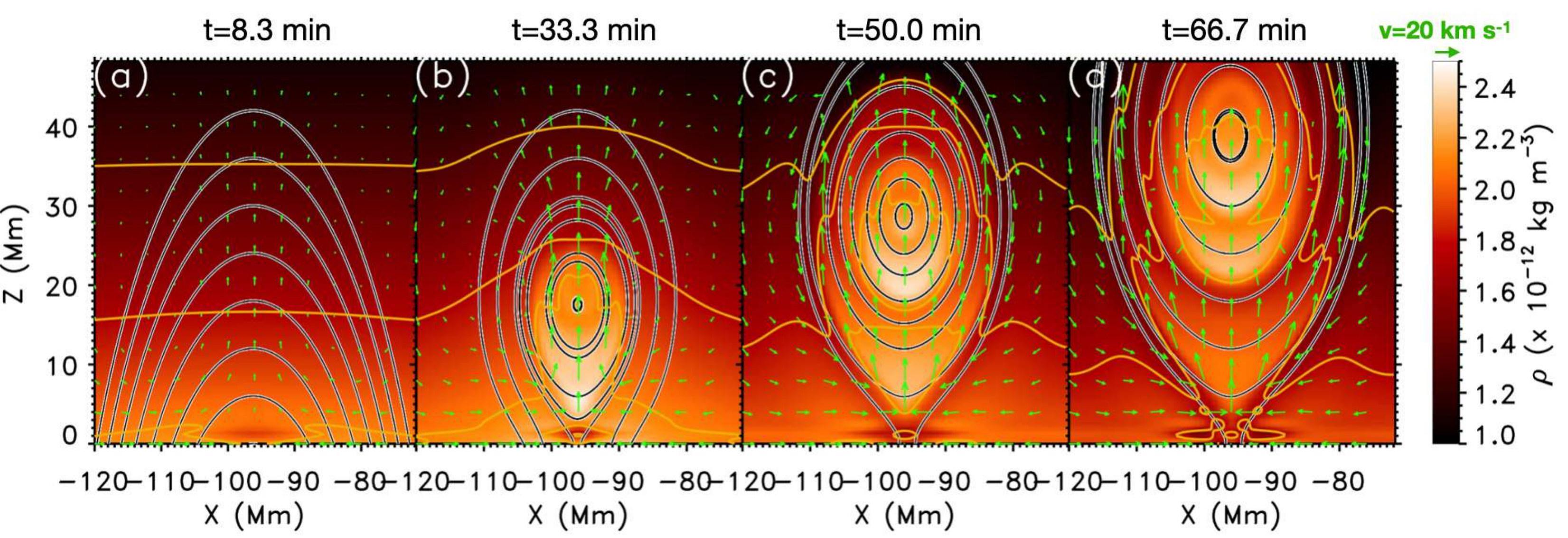}
	\caption{Evolution of the density and the magnetic field lines during FR1 formation. The yellow lines denote the density isocontours. The green arrows show the velocity field. Note that field lines in the different panels are not the same field lines carrying the same plasma. They are drawn to reflect the topological structure of the magnetic field at a given time. \label{fig:eruption-formation-1}}
\end{figure*}

We use the \citet{vanBallegooijen:1989apj} mechanism to form flux ropes from the sheared arcades with the footpoints motions. For FR1, the horizontal velocity imposed at the base in the direction towards the PIL (see green arrows in Fig. \ref{fig:eruption-formation-1}) is given by
\begin{equation}\label{imposed_velocity_vx_1}
	V_{x}(x, t)=-V_{0}(t)\sin{\left[\frac{2\pi(x-x_{c})}{W}\right]}\exp{\left[-\frac{(x-x_{c})^2}{2\sigma^2}\right]} \, ,
\end{equation}
where $W=576.0 \Mm$ defines the horizontal extension of the domain, $x_c=-96 \Mm$ is the central position of the function centered at the corresponding PIL, and $\sigma=13.6 \Mm$ is the half-width of the converging region. We additionally set the $V_y$ and $V_z$ velocity components to zero at the bottom of the domain. 
The activation and deactivation of the convergence are controlled by a function $V_0(t)$ defined as
\begin{eqnarray}\nonumber
	V_0(t)= V_{conv}  \left\{ 0.5\left[\erf\left(\frac{t-2\lambda-t_1}{\lambda}\right)+1\right] \right. \\ \label{temp_evolution-1}
	\left.-0.5\left[\erf\left(\frac{t-2\lambda-t_2}{\lambda}\right)+1\right] \right\}  \, ,
\end{eqnarray}
where $V_{conv}=12\kms$  is the maximum converging velocity. The flux rope is formed during the time defined by the parameters $t_1=100\secs$ and $t_2=4000\secs$. The parameter $\lambda=150\secs$ is responsible for the smoothness during the activation and deactivation of the converging process.
As explained in \citet{Kaneko:2015apj}, the temperature increases inside the flux rope due to the magnetic reconnection. This leads to an increase in the gas pressure and, consequently, of the plasma-$\beta$. In order to avoid this heating of the plasma inside FR1, we use Newton's cooling law given by
\begin{equation}\label{eq:newtonian-heat}
	Q_\mathrm{R}=-c_v\rho_0 \, \frac{T-T_0}{\tau_\mathrm{R}} \, ,
\end{equation}
where $T$ and $T_0$ are the temperature at any time and the initial temperature, respectively, $\rho_0$ is the initial density, $\tau_\mathrm{R}$ is the radiative relaxation time equal to $3\secs$ everywhere in the domain, and $c_v$ is the specific heat at constant volume. We use radiative cooling only during the formation of the FR1 up to the time $t=4000 \secs$. 
This heat loss term allows us to keep the temperature of the system always close to the initial temperature, i.e., $1.5$ MK, everywhere in the domain.

Different stages of the formation are shown in Fig. \ref{fig:eruption-formation-1}. Panel  \ref{fig:eruption-formation-1}a shows the initial field of the magnetic arcade. Panel \ref{fig:eruption-formation-1}b shows how the foot points of the field lines move towards $x=-96\Mm$ following Eq. \eqref{imposed_velocity_vx_1}. When the field lines approach each other, they reconnect, forming a twisted flux rope. As we continue the converging process, the flux rope slowly rises to the higher positions (Figs. \ref{fig:eruption-formation-1}c and d). In order to prevent the eruption of the FR1, we stopped the convergence at $t=66.7\min$, when the center of the flux rope is located at the height of $39 \Mm$ (see Fig. \ref{fig:eruption-formation-1}d). The magnetic field strength in the dipped region ranges between $7-9$ G. Figures \ref{fig:eruption-formation-1}b-d show that the rising flux rope has a slightly increased density inside the dipped region which implies that the flux rope lifts some plasma from the lower atmosphere. In recent 2.5D numerical studies, \citet{jenkins_prominence_2021} and \cite{Brughmans:2022aap} showed in more detail the process of prominence formation in the levitation-condensation models. Unlike our case, their model included background heating and thermal conduction, which, in addition to radiative losses, allowed thermal instability to develop. As the instability develops, the prominence is formed in situ by condensing the lifted plasma inside the flux rope. In our adiabatic experiment, when the magnetic rope is formed, it also lifts a small amount of plasma from the bottom of the domain. However, it does not condense, and we produce the prominence loading by an artificial mechanism described below.

As mentioned before, in this experiment, the eruption of a second flux rope, FR2, is the main trigger of the oscillations. 
As we aim to produce an energetic eruption, we included the shearing motions in addition to the converging ones. This increases the shear angle and, consequently, the axial component of the magnetic field $B_y$. The converging velocity, $V_{x}$ at the bottom of the domain is defined by Eq. \eqref{imposed_velocity_vx_1} with $W=576 \Mm$, $x_c=0 \Mm$, central position located at the corresponding PIL, $\sigma=13.6 \Mm$, while $V_y=-V_x$. The vertical $V_z$ velocity component remains zero at the bottom of the domain. 
Temporal evolution is defined by Eq. \eqref{temp_evolution-1} with the following parameters:
 $V_{max}$ is either the maximum converging velocity $V_{conv}=4 \kms$, or maximum shearing velocity, $V_{shear}=18\kms$. The rest of the parameters are $t_3=4000\secs$,  $t_4=10000\secs$, and $\lambda=150\secs$. The FR2 is formed in the absence of radiative losses.
\begin{figure*}[!ht]
	\centering\includegraphics[width=0.85\textwidth]{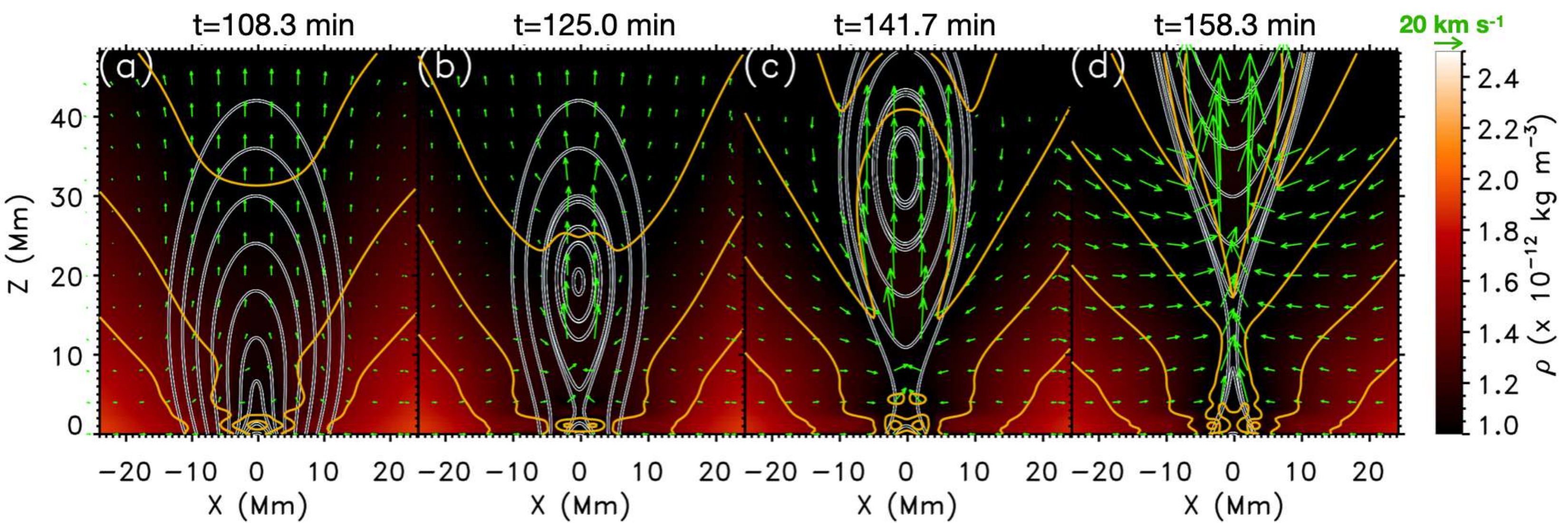}
	\caption{Evolution of the density and the magnetic field lines during the FR2 formation due to the converging and shearing motions at the base. The yellow lines denote the density isocontours. The green arrows show the velocity field. Note that field lines in the different panels are not the same field lines carrying the same plasma. They are drawn to reflect the topological structure of the magnetic field at a given time. \label{fig:eruption-formation-2}}
\end{figure*}
Figure \ref{fig:eruption-formation-2} shows the formation process of the FR2 after the formation of the FR1. It can be noted that the evolution slightly differs from the evolution of the FR1. Since we use the shearing motions together with the converging ones, the shear angle of the magnetic structure increases. Figure \ref{fig:eruption-formation-2}b shows that the FR2 has more elliptical magnetic field lines in the 2D projection than in the case of the FR1, related to the increase of the axial magnetic field. In contrast to FR1, FR2 does not have a large amount of dense plasma in the bottom helical part. Since the FR2 formation is dominated by the shearing motions, it seems that, in this case, the forming flux rope collects and lifts less plasma.
From Fig. \ref{fig:eruption-formation-2}d, we can see the onset of the flux rope eruption at around $t=158$ minutes. The velocity field shows a large upward motion in the flux rope associated with its rise. 
We used symmetric (for $V_{x}, V_{y}, B_{z}, p, \rho, T, e_{int}, e$) or antisymmetric (for $V_{z}, B_{x}, B_{y}$) boundary conditions at the top boundary. The system is assumed to be periodic at the side boundaries. At the bottom, the magnetic field evolves according to the velocity field during the flux ropes formation process. Later on, the line-tied conditions are applied using zero velocities. The rest of the variables are assumed to be symmetric at the bottom boundary. 

After the formation of FR1, we load a prominence in the dips of the flux rope, adding a source term in the continuity equation as explained in \citet{Liakh:2020aap, Liakh:2021aap}. The source term is defined as:
\begin{equation}\label{eruption_mass_source}
	S_{\rho}=\frac{\chi\rho_{0}}{t_{\rm load}}\exp\left({-\frac{(x-x_p)^4}{\sigma_x^4}-\frac{(z-z_p)^4}{\sigma_z^4}}\right) \, .
\end{equation}
The parameters $\rho_{0}$ and $\chi$ are the background density and the density contrast, respectively. The mass is loaded during the time $t_{3}=4000\secs$ and $t_{4}=4100\secs$ and the time parameter $t_{load}=t_{4}-t_{3}=100\secs$ is associated with the loading rate. The spatial parameters $x_p=-96 \Mm, z_p=25.4 \Mm$, are coordinates of the center of the distribution, and $\sigma_x=3.6 \Mm, \sigma_z=4.8 \Mm$ are related to the half-size of the mass distribution region.
In all the numerical experiments considered in this paper, we used a reduced value of the density contrast, $\chi=30$, with respect to the previous numerical studies by \citet{Zhou:2018apj}, \citet{Adrover:2020aap}, and \citet{Liakh:2020aap, Liakh:2021aap}  who used values between 50 and 200. In the case of the flux rope configurations, we choose a lower density contrast in order to avoid problems in the prominence support and to avoid a strong dropping down of the flux ropes as in \citet{Hillier:2013apj}. Using a lower density contrast in the dipped arcade configuration, we expect a smoother mass loading and less perturbation of the magnetic field by this loading process. Thereby, a shorter relaxation time can be applied to remove those velocities compared to \citet{Adrover:2020aap}.
\begin{figure*}[!ht]
	\centering\includegraphics[width=0.85\textwidth]{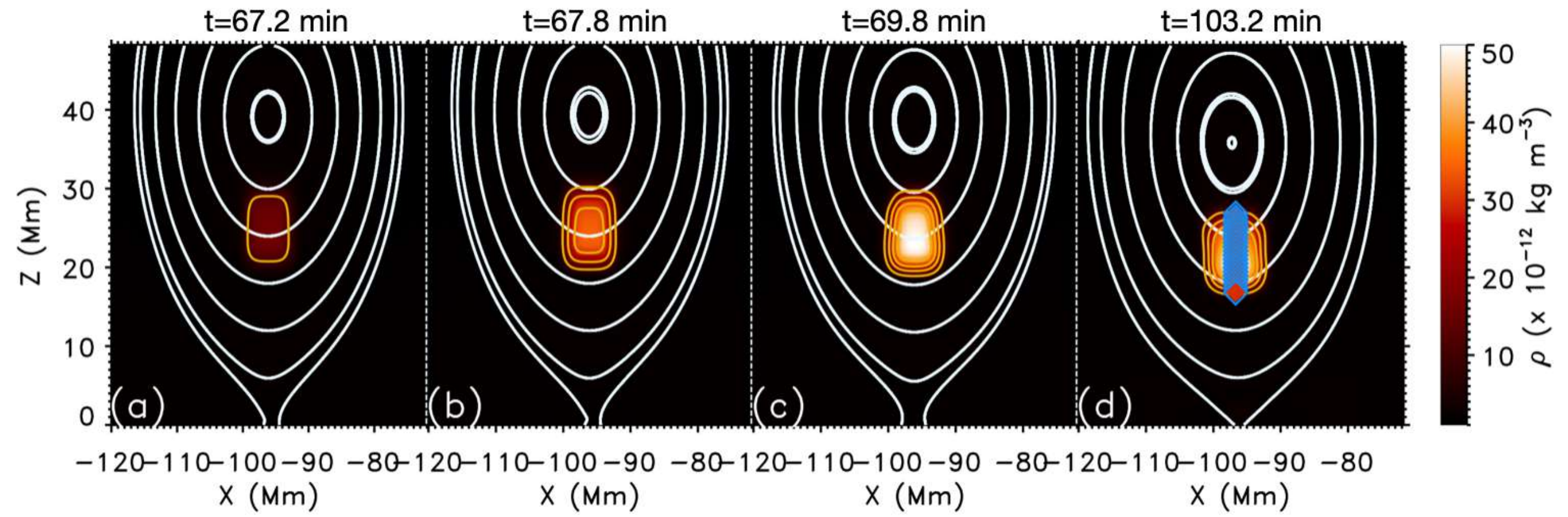}
	\caption{Evolution of the density and the magnetic field lines during and after the mass loading process. The yellow lines denote the density isocontours. In panel d, the sequence of the red-blue diamonds shows the positions of the fluid elements used for the analysis of motions. Note that field lines in the different panels are not the same field lines carrying the same plasma. They are drawn to reflect the topological structure of the magnetic field at a given time. \label{fig:eruption_mass_loading}}
\end{figure*}
Figures \ref{fig:eruption_mass_loading}a-c show that during the mass loading process, the flux rope center drops by a few Mm to find a new equilibrium position. The magnetic configuration accommodates the heavy mass against the gravity in the dipped region. After mass loading, the system evolves for 16.7 minutes, freely settling in a quasi-equilibrium (Fig. \ref{fig:eruption_mass_loading}d) with small vertical oscillations.

\subsection{Evolution of the eruption and of the prominence}
\begin{figure*}[!ht]
	\centering\includegraphics[width=0.95\textwidth]{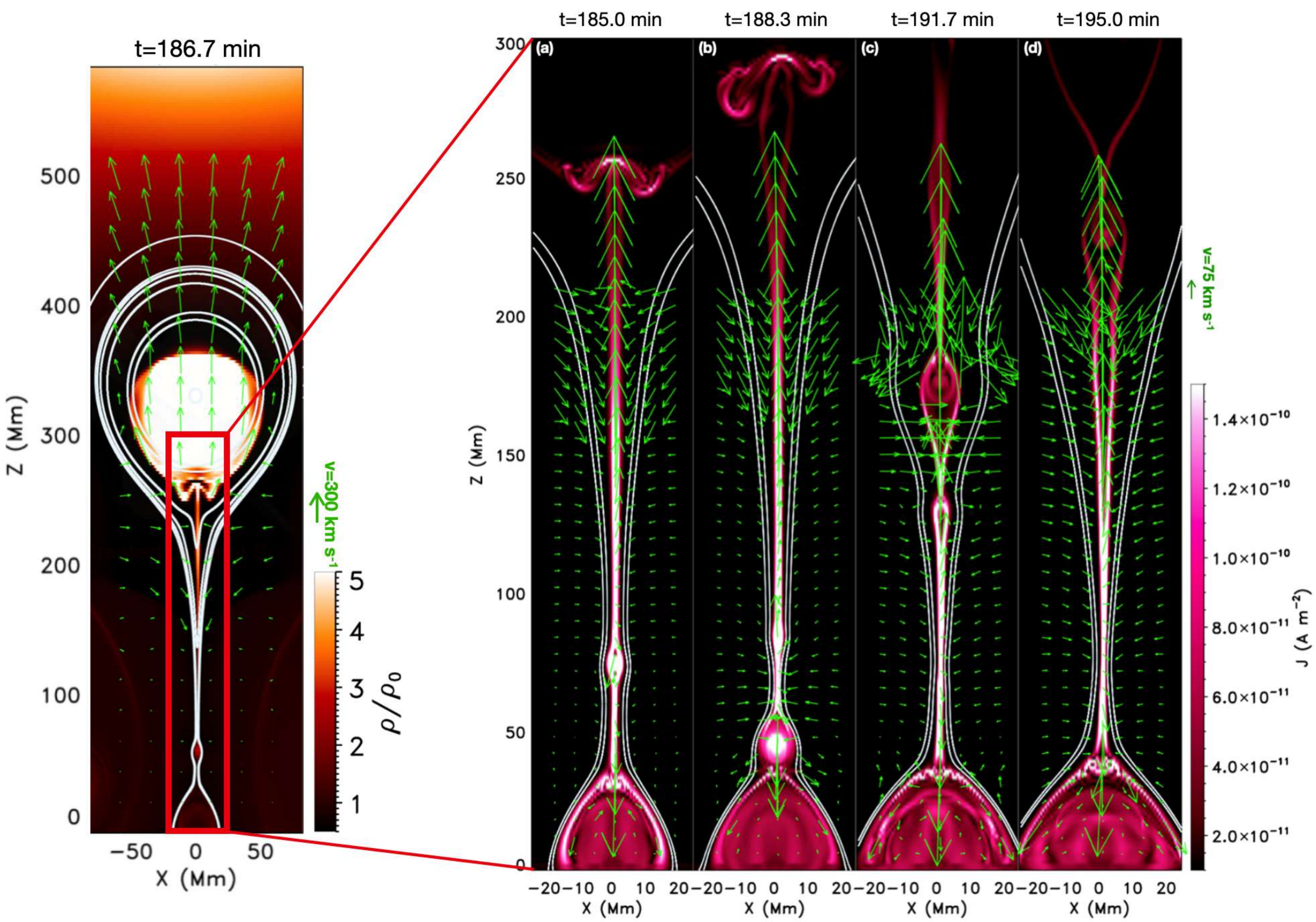}
	\caption{Snapshot of the density contrast and the magnetic field lines during the eruption of the FR2 (left panel) and the current density distribution (color) and magnetic field lines (white lines) underneath erupting FR2 (right panel). The green arrows denote the velocity field. The red box in the left panel delimits the region shown in the right panels. \label{fig:eruption_evolution_current}}
\end{figure*}

When we stop the shearing and converging motions around $x=0\Mm$ at $t=166.7\mins$, FR2 has already reached the critical height where its loss of equilibrium and eruption are unavoidable. Therefore, FR2 continues moving upwards even when we deactivate the velocities at the foot points. Figure \ref{fig:eruption_evolution_current} shows the rising flux rope. From the velocity field shown by the green lines in the left panel of Fig. \ref{fig:eruption_evolution_current}, we find that the flux rope strongly accelerates during its rise. At the time of $186.7\mins$, the velocity of the eruption is around $300\kms$. The density contrast shows that the flux rope carries inside the denser plasma compared to the surrounding. We note a reduced density region, the so-called dimming region \citep{Moses:1997solphys,Thompson:1998grl} above the flux rope at height from $z=400$ to $450\Mm$. Slightly above the dimming region ($z>550\Mm$), we can see an enhanced density area formed as the flux rope rises by pushing the plasma above it. By comparing this behavior with other numerical works, we know that this region of the enhanced density is associated with the propagation of an EIT front \citep[see, e.g.,][]{Chen:2002apjl}. The white lines show that a vertical current sheet is formed below the flux rope. The current sheet is fragmented, and one plasmoid is seen (Fig. \ref{fig:eruption_evolution_current}a). This plasmoid is a mini flux rope and has an increased density contrast. We can compare our study with the work by \citet{Zhao:2017apj,Zhao:2019apj}, who studied the properties of the standard solar flare model using non-adiabatic 2.5D numerical simulations. \citet{Zhao:2017apj,Zhao:2019apj} have shown that the plasmoid formation leads to a decrease in the density of the current sheet, suggesting that the plasmoids transport dense material upwards and downwards along the current sheet. 
 In our experiment, we observe an identical behavior with plasmoids moving in both directions in the current sheet resulting in a decrease in the density along this current. Upward-traveling plasmoids merge with the magnetic rope (Figs. \ref{fig:eruption_evolution_current}c and d). Figure \ref{fig:eruption_evolution_current} also shows that under the current layer, there is an arcade of magnetic loops. These structures are formed due to magnetic reconnection and are usually called post-reconnection loops. Downward-traveling plasmoids ultimately collide with this arcade (Figs. \ref{fig:eruption_evolution_current}a and b).

\begin{figure*}[!ht]
	\centering
	\includegraphics[width=0.97\textwidth]{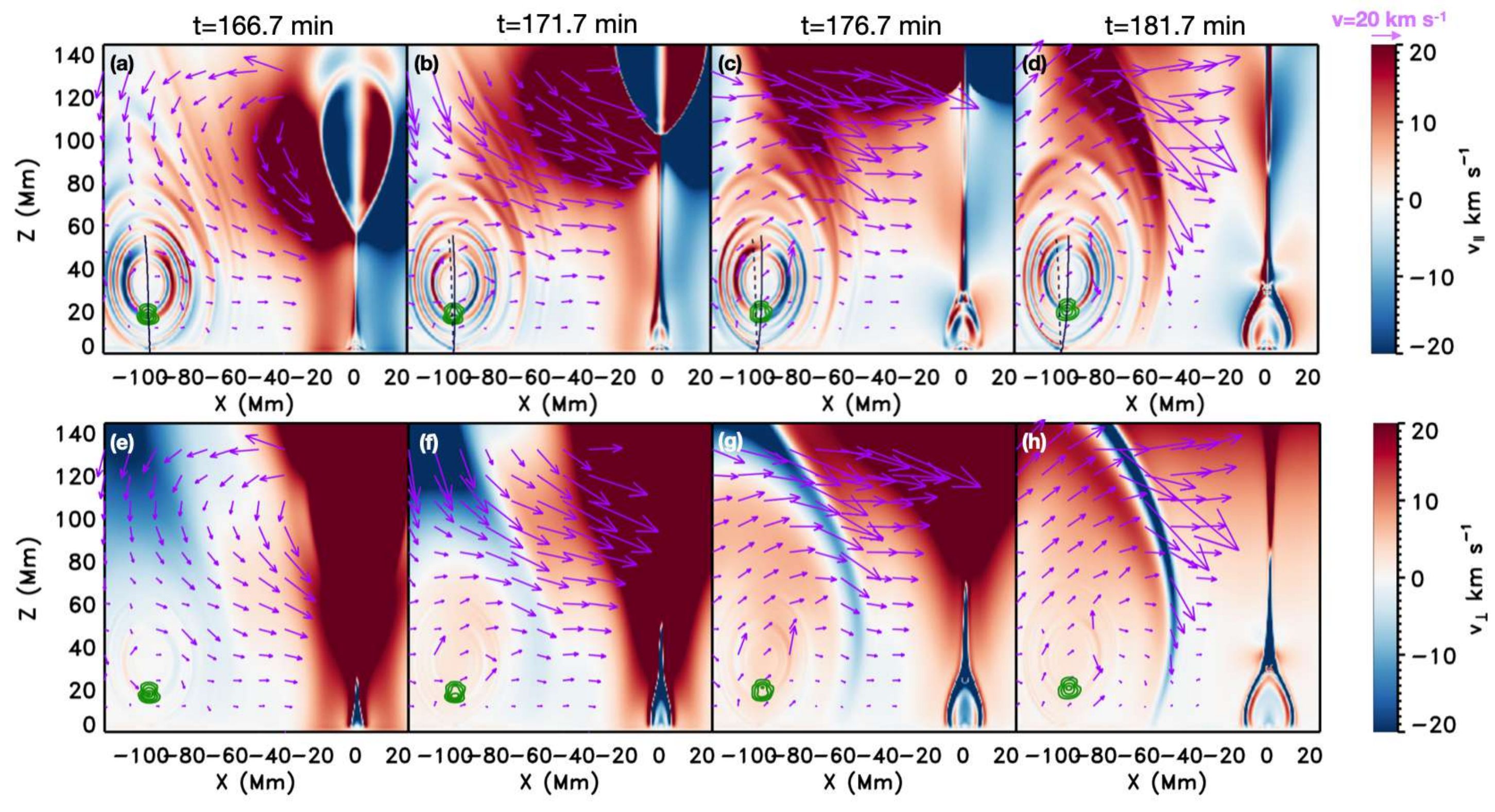}
	\caption{Evolution of $v_{\parallel}$ (top panel) and $v_{\perp}$ (bottom panel) in a region around the FR1 during the FR2 eruption (saturated). The purple arrows denote the velocity field. The green lines show the density isocontours. The dashed and solid black lines denote the vertical axis of the flux rope defined by the criterion $B_z=0$ at time $166.7\mins$ and time shown in the corresponding panels, respectively. \label{fig:eruption_velocity_initial}}
\end{figure*}


Figures \ref{fig:eruption_velocity_initial}a-d show the evolution of the velocity component along the magnetic field, $v_\parallel$, defined as
\begin{eqnarray}
	v_{\parallel}&=&\frac{(\mathbf{v} \cdot \mathbf{B})}{|B|}\, , \label{long-velocity}
\end{eqnarray}
in a region that includes both the erupting and prominence-hosting flux ropes. The time sequence is taken during the onset of the eruption. Figure \ref{fig:eruption_velocity_initial}a shows when the center of the flux rope reached approximately $100\Mm$ height. The eruption produces velocity disturbances around it. Above the FR2, we see the diverging flow, implying the rarefaction of the plasma (Fig. \ref{fig:eruption_evolution_current}, right panel). 
Right below the FR2, the situation is the opposite. Instead of diverging flow, strong converging motions can be noted around the $z$-axis. The diverging flow above and the converging flow below the FR2 lead to the counterclockwise flow at the left side of the flux rope (see purple arrows in  Fig. \ref{fig:eruption_velocity_initial}a). When the FR2 reaches higher positions, the influence of the flows around it decreases (Figs. \ref{fig:eruption_velocity_initial}b-d). However, a converging flow remains directed to the current sheet (Fig. \ref{fig:eruption_velocity_initial}). 
Comparing the position of the dashed and solid lines in Figs. \ref{fig:eruption_velocity_initial}a-d, we note that the FR1 gradually becomes more inclined toward the current sheet due to the influence of the converging flow.

Figures \ref{fig:eruption_velocity_initial}e-h show the transverse velocity, $v_\perp$, defined as
\begin{eqnarray}
	v_{\perp}&=&\left(\mathbf{v} - v_{\parallel} \frac{ \mathbf{B}}{B}\right) \cdot \mathbf{\hat{e}_{z}} \, , \label{trans-velocity}
\end{eqnarray}
where $\mathbf{\hat{e}_{z}}$ is the unit-vector along the $z$-axis, during the propagation of the eruption.  The very red region corresponds to the largest velocity of the upward propagation of the FR2. 
We do not detect any coronal Moreton or EIT wave propagating toward the FR1.
We consider several possibilities for why these waves are not produced in our simulations. The first one is that in our experiment, the typical value of the magnetic field strength is $10$ G, implying that the magnetic energy can be insufficient to produce an intense wave. 
Another reason can be associated with the height at which this wave front develops. This height depends on the height of the loss of equilibrium of the flux rope. For instance, \citet{Zhao:2022apj} obtained the wave front forming very low in the atmosphere (below $50$ Mm). Forming such a small-scale front was possible in their case because the configuration was initially in non-equilibrium. In our experiment, the loss of the equilibrium happens very high in the atmosphere due to the large magnetic scale height $H_{B}=118$ Mm, and, consequently, the fast-mode wave front develops at larger distances from the region $z<140\Mm$ shown in Fig. \ref{fig:eruption-initial}, and, consequently, does not affect the FR1 and prominence.
Finally, the fast-mode wave front never becomes a shock front in the horizontal direction, unlike in the vertical direction. The reason is that the magnetic field remains strong in the low corona, even further away from the reconnection site. This means that an ordinary fast magnetoacoustic wave does not become a shock. Unlike our case, \citet{Zhao:2022apj} used a quadrupolar magnetic field structure in which strength decreases away from the eruption site. As a consequence, the fast-mode wave becomes a dome-like shock wave.
The EIT wave-like phenomenon neither appears in Fig. \ref{fig:eruption_velocity_initial}. As we see from the left panel of Fig. \ref{fig:eruption_evolution_current}, the density enhancement associated with the EIT front also appears much higher in the corona ($z>500$ Mm). Therefore EIT front cannot affect the neighboring flux-rope prominence at height $\sim20\Mm$.

The right panels of Fig. \ref{fig:eruption_evolution_current} represent a closer look at the current sheet during the flux rope rise. Figures \ref{fig:eruption_evolution_current}a-d show that more field lines are reconnected during the process, and the region of the post-reconnection loops expands at both sides of the PIL while the current sheet elongates and becomes fragmented due to the appearance of the plasmoids after $180\mins$. One of the early-formed plasmoids forms at the height of $60-70\Mm$ and starts to propagate downwards toward the loops (Figs. \ref{fig:eruption_evolution_current}a and b). 
The velocity field shows that the plasmoid approaches the post-reconnection loops at the bottom having a large velocity. Plasmoid-loops confluence also causes large velocities around the interaction region (see green arrows). In this way, the plasmoids can produce additional disturbances affecting the surroundings, including the nearby flux rope prominence.
At later times Figs. \ref{fig:eruption_evolution_current}c and d show the propagation of the plasmoids in the direction to the erupting flux rope. According to \citet{Zhao:2022apj}, upward-propagating plasmoids feed a bottom helical part of the flux rope with a dense plasma, leading to the formation of the prominence threads under thermal instability during the flux rope rising phase. In our adiabatic case, this merging simply increases density inside FR2.
Later, the plasmoids are chaotically formed in the current sheet and propagate both upwards and downwards.

\begin{figure*}[!ht]
	\centering\includegraphics[width=0.97\textwidth]{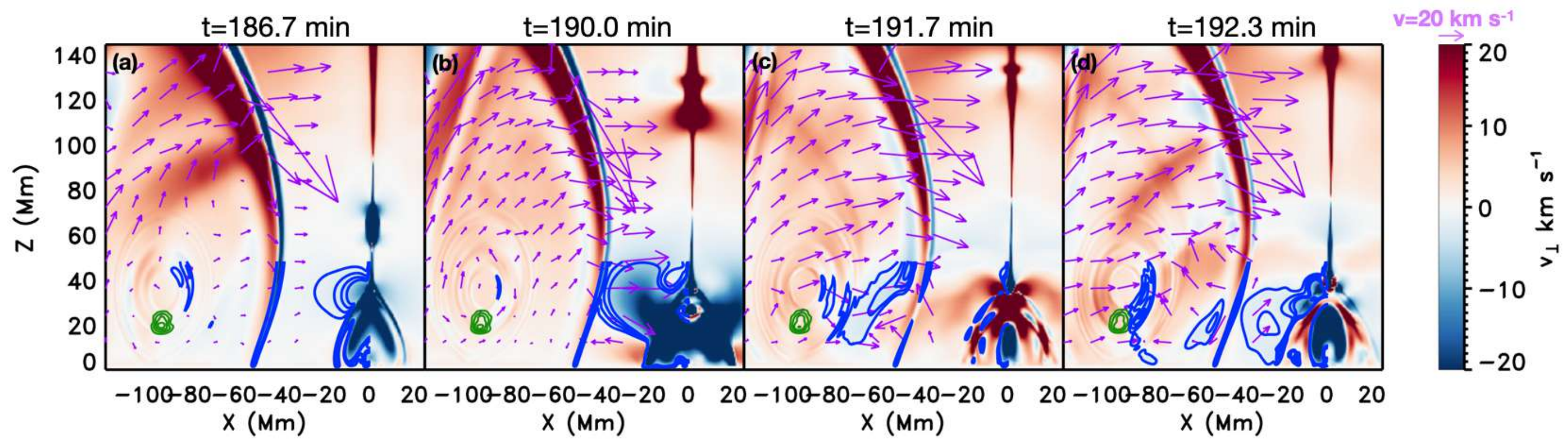}
	\caption{Evolution of $v_{\perp}$ in a region around the FR1 during the plasmoids formation in the current sheet of the FR2 (saturated). The plasmoid is seen in panel a, at height $50-70\Mm$, in the elongated current sheet. The purple arrows denote the velocity field. The green lines show the density isocontours. The blue lines denote $v_{\perp}$ isocontours. \label{fig:eruption_velocity_plasmoid}}
\end{figure*}
We aim to understand the perturbations created by the plasmoids and whether those disturbances are able to propagate across the magnetic field, reaching distant regions such as FR1. Figure \ref{fig:eruption_velocity_plasmoid}a shows a recently formed plasmoid that starts to move downwards. When it approaches the loops at the bottom, it creates a perturbation seen in Fig. \ref{fig:eruption_velocity_plasmoid}b. In order to follow the wave front created by this perturbation, we added velocity isocontours (see the blue lines). We can observe this in Figs. \ref{fig:eruption_velocity_plasmoid}a and b that the left part of the wave front moves toward the FR1 and the prominence. Figure \ref{fig:eruption_velocity_plasmoid}c shows that the disturbance front enters the region of the FR1, and, at $t=192.3\mins$, it reaches the prominence mass (Fig. \ref{fig:eruption_velocity_plasmoid}d). Another plasmoid located higher in the current sheet is shown in Figs. \ref{fig:eruption_velocity_plasmoid}b-d. This plasmoid propagates upwards, and its motion does not seem to affect the FR1.

\begin{figure*}[!ht]
	\centering\includegraphics[width=0.97\textwidth]{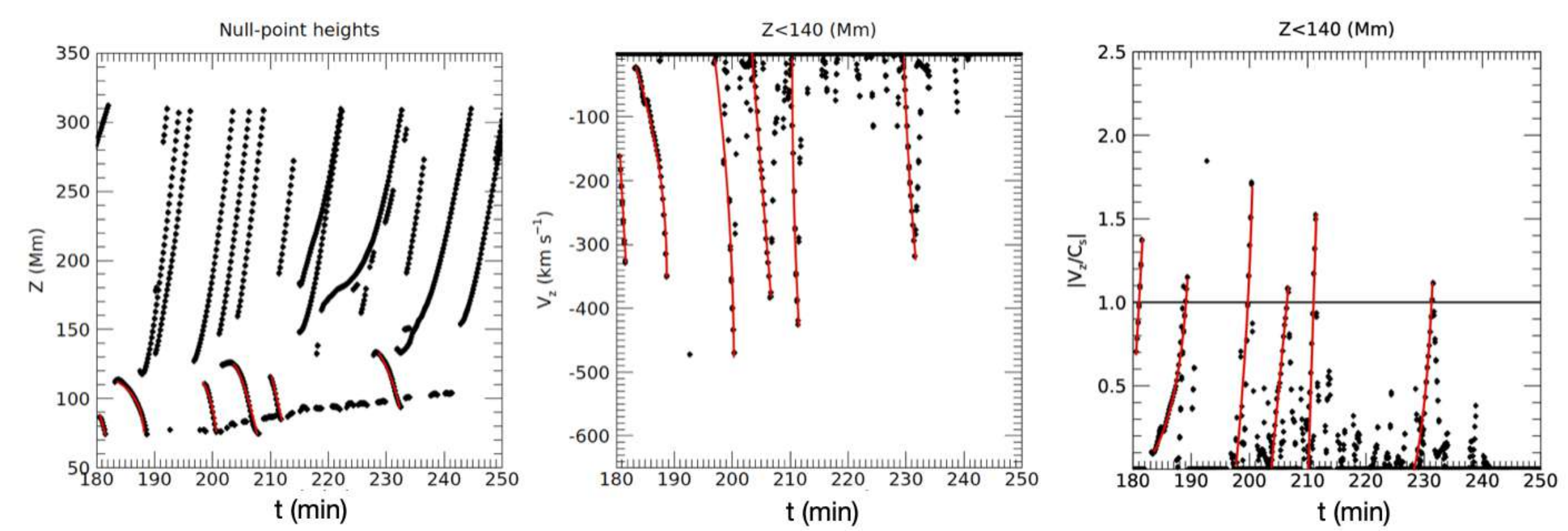}
	\caption{Temporal evolution of the detected null-points. Panels from left to right: height vs. time of nulls during the current sheet fragmentation; $v_{z}$ of the plasmoids formed below $Z=140$ Mm; downward velocity to the local sound speed ratio for plasmoids formed below $Z=140$ Mm. The red lines denote the plasmoids which merge with post-reconnection loops. \label{fig:eruption_plasmoids}}
\end{figure*}

According to Fig. \ref{fig:eruption_velocity_plasmoid}, only plasmoids moving downwards are able to produce perturbations in the prominence. Therefore, it is interesting to have a closer look at the dynamics of the plasmoids with descending motions. At the center of the plasmoid, there is a null-point of the magnetic field. Thus, we define a criterion to detect these points by finding the sign change of $B_x$ along the $X=0$ Mm axis. This way, we measure the $Z-$coordinate of the null points, their vertical velocities, $v_{z}$, and the sound speed at the exact location. The results are shown in Fig. \ref{fig:eruption_plasmoids}. The left panel shows the height of the null points as a function of time. We can see that the first null points that appear in the current sheet at low heights move downwards. However, at later times, the majority of them move upwards.  
\citet{Karpen:2012apj} obtained a similar result from their 2.5D simulations, i.e., downward-moving plasmoids dominating the first minutes of fragmentation of the current sheet, and the situation is opposite later. From Fig. \ref{fig:eruption_plasmoids}(left panel), we see that the downward-moving plasmoids are formed approximately below the height of $140$ Mm.
The downward velocity, $v_{z}$, and ratio, $v_{z}/c_{S}$, for those plasmoids are displayed at the central and right panels of Fig. \ref{fig:eruption_plasmoids} respectively. 
The plasmoids form and rapidly reach a speed of several hundred $\kms$.
For instance, the plasmoid around $t= 200\mins$ has a velocity around $400\kms$ when reaching the top part of the post-reconnection loops. 
The velocities acquired by the plasmoids can be supersonic. Therefore, these plasmoids form shocks when merging with the post-reconnection loops.

\begin{figure*}[!ht]
	\centering
	\includegraphics[width=0.97\textwidth]{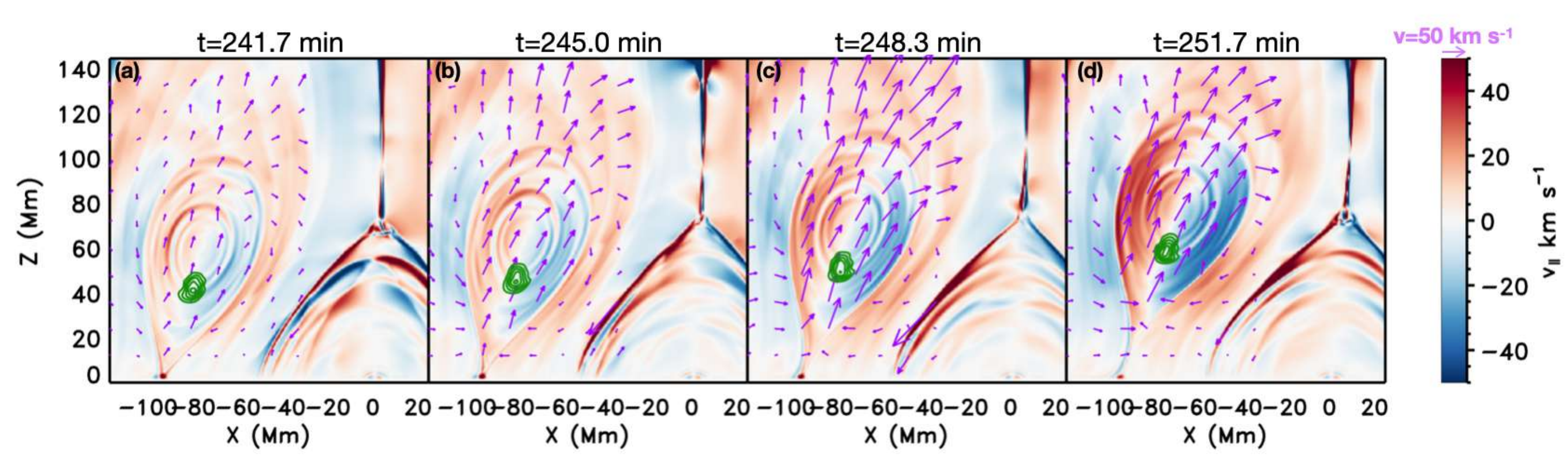}
	\caption{Evolution of $v_{\parallel}$ during the eruption of the FR1 (saturated). The purple arrows denote the velocity field. The green lines show the density isocontours. \label{fig:eruption_velocity_secondary}}
\end{figure*}

The converging flow toward the current sheet continuously affects the FR1 (green and purple arrows in Figs. \ref{fig:eruption_evolution_current} and \ref{fig:eruption_velocity_initial}). This disturbance pulls FR1 up and to the right. At the time of $241.7\mins$, FR1 becomes unstable and starts to rise in the direction to the current sheet. Figs. \ref{fig:eruption_velocity_secondary}a-d, show the evolution of the longitudinal velocity. One can observe converging flow below the flux rope and diverging flow above it. Such a behavior resembles the onset of the eruption of the FR2, although due to the inclination of the FR1, the diverging region is not so symmetric with respect to the flux rope axis. Later on, the FR1 continues moving upwards and to the right. Figure \ref{fig:eruption_velocity_secondary}d shows that the current sheet starts to form below FR1. During the time interval shown in this figure, FR1 is greatly accelerated upwards, and its center reaches $\sim80\Mm$. 
Since the FR1 hosts heavy prominence plasma, an enhanced gravity force acts downward. Moreover, this prominence-hosting flux rope was formed solely by converging motions, not significantly affecting the axial component of the magnetic field. Consequently, the magnetic pressure force, considered the main driving force of the eruption, can be insufficient compared with the downward magnetic tension force of the surrounding arcades. Altogether, such force imbalance leads to a failed eruption of the FR1. The FR1 reaches a maximum height of around $160$ Mm and then falls back to the bottom of the numerical domain.

\subsection{Prominence oscillations}\label{sec:prominence-oscillations-eruption}

We are interested in studying the plasma oscillatory dynamics in the different prominence regions and the longitudinal and transverse directions with respect to the magnetic field. It is convenient to use two approaches. The first one is to trace the magnetic field lines and compute the rest of the magnitudes along these field lines. The second approach is to advect the velocity field following a set of fluid elements, hereinafter referred to as corks. The first approach is used when studying the plasma motions in a quasi-static magnetic field, which implies slow temporal evolution. This method will be used for the analysis in Sects. \ref{sec:prominence-oscillations-flux-ropes} and \ref{sec:dipped-arcade}. As shown before, FR1 is significantly influenced by the dynamics of FR2. These dynamics create significant changes in its magnetic field structure as inclination or eruption of FR1. We, therefore, use the second approach computing the magnitudes along the traces of the particles. The initial positions of the corresponding corks are shown in Fig. \ref{fig:eruption_mass_loading}d. The plasma elements have the initial coordinates from $(x,z)=(-96,16.8)$ to $(-96,26.8)\Mm$. We start to follow the fluid elements when FR2 erupts at $t=166.7\mins$.
\begin{figure*}[!ht]
	\centering
	\includegraphics[width=0.9\textwidth]{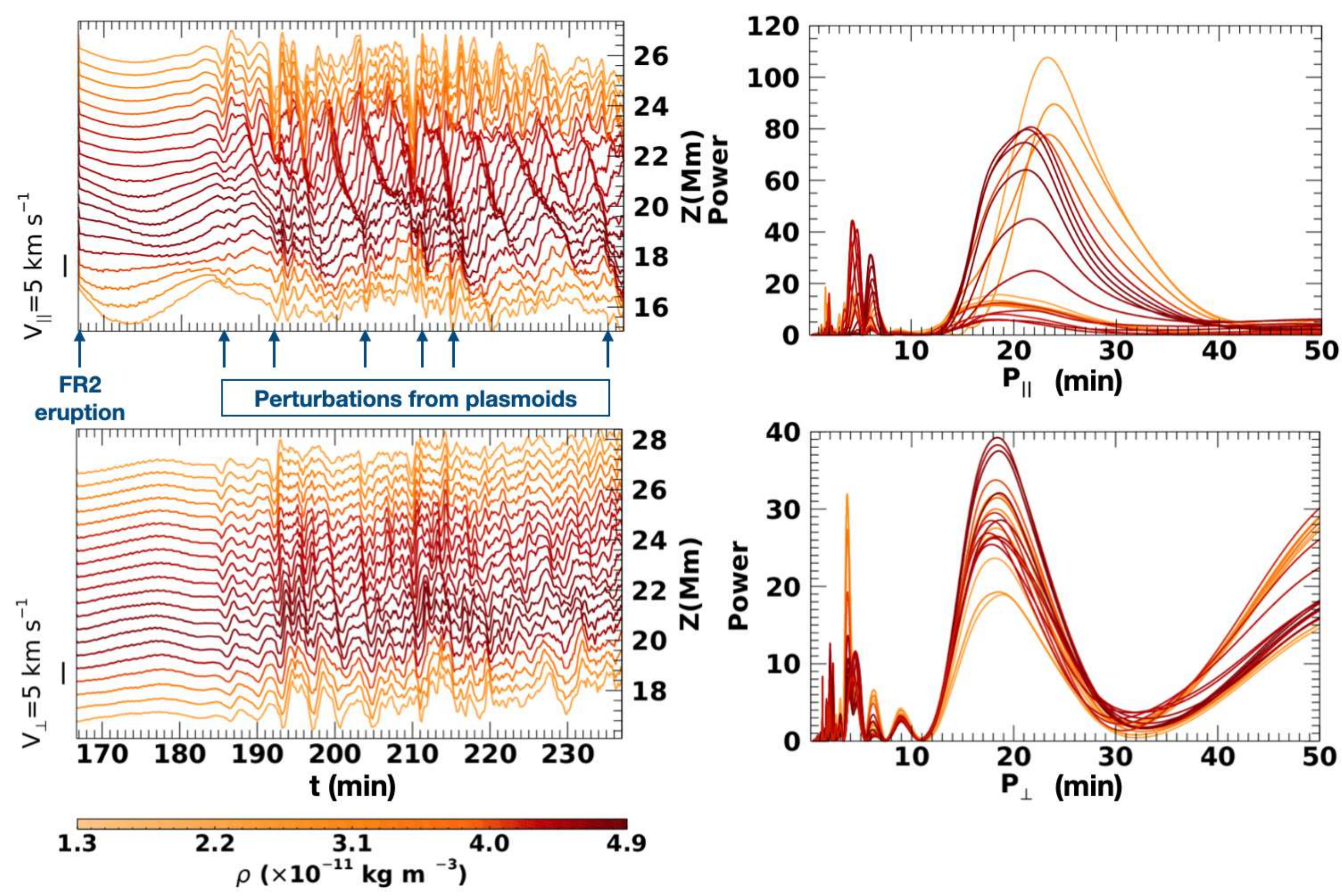}
	\caption{Temporal evolution of the velocity field in the prominence region. Left panels: $v_{\parallel}$ (top) and $v_{\perp}$ (bottom) of the fluid elements shown with red-blue diamonds in Fig. \ref{fig:eruption_mass_loading}d after the eruption of FR2. The blue arrows denote the time arrival of the perturbation from the confluence of the plasmoids denoted with the red lines in Fig. \ref{fig:eruption_plasmoids}. Right panels: the periodograms of the $v_{\parallel}$ (top) and $v_{\perp}$ (bottom) in time-interval
		$166.7$-$233\mins$. The color bar denotes the density of the fluid elements. The left vertical axis indicates the velocity amplitude scale. The right vertical axis denotes the initial vertical positions of the fluid elements.\label{fig:eruption_vpar_vper_corks}}
\end{figure*}

In Fig. \ref{fig:eruption_vpar_vper_corks}, we note a small variation of both velocity components, not exceeding $5 \kms$ in the denser part of the prominence during the first phase up to $185\mins$. These variations of the longitudinal and transverse velocity occur due to the inclination of FR1. A first significant velocity peak appears at $t=185\mins$ (left panels of Fig. \ref{fig:eruption_vpar_vper_corks}), coinciding in time with the arrival of the perturbation caused by the first plasmoid in the current sheet. After that, multiple peaks can be seen in the signals denoted by the blue arrows. Remarkably, the plasmoids produce stronger disturbances of prominence compared to the actual eruption of the FR2. The induced oscillations, however, are SAOs rather than LAOs. Their typical velocity amplitude is around $5\kms$.
The longitudinal velocity, $v_\parallel$, combines different oscillatory modes. Furthermore, the long-period oscillations seem to be not coherent with height. The periodogram obtained for $v_{\parallel}$ using the Lomb-Scargle algorithm \citep{Lomb:1976apss,Carbonell:1991aap} for the time-interval $185$-$237\mins$, is displayed at the upper right panel. The longitudinal oscillations are dominated by long-period mode with periods ranging between $18.0-23.3\mins$. This period tends to decrease with height. We further discuss the long-period longitudinal oscillations in Sect. \ref{sec:prominence-oscillations-flux-ropes}. The periodogram also shows peaks corresponding to the two short-period modes in the range of $4.4-4.6\mins$ and $6.2-6.4\mins$. 

Vertical SAOs in $v_\perp$ are present at every field line (bottom left panel of Fig. \ref{fig:eruption_vpar_vper_corks}). Their dynamics are complex, with a combination of periodicities reflected in the corresponding periodogram (right panel). Vertical oscillations are more synchronized with height than longitudinal oscillations. Starting from $200\mins$, the average velocity of all the fluid elements increases gradually. This velocity increase is associated with the slow rise phase of the FR1 (Fig. \ref{fig:eruption_velocity_secondary}) when due to the effect of the converging flow around FR2, FR1 loses equilibrium and erupts. The periodogram shows that the dominant period has values in the range of $17.5-18.7\mins$ for the different field lines. Other peaks in the periodogram correspond to the short-period oscillations. In the densest prominence region, we distinguish two main periodicities $3.7-4.5\mins$ and $1.3-2.3\mins$. The short-period vertical oscillations are further discussed in Sect. \ref{sec:prominence-oscillations-flux-ropes}. The right panels of Fig. \ref{fig:eruption_vpar_vper_corks} show that some oscillatory modes are present in both longitudinal and transverse velocity variations. For instance, long period, $\sim20$, and short period, $\sim4$, peaks are present in both periodograms, suggesting the longitudinal to transverse modes coupling. 

\section{External perturbation in association with two distant flux rope prominences}\label{sec:two-prominences}

\begin{figure}[!t]
	\centering
	\includegraphics[width=0.5\textwidth]{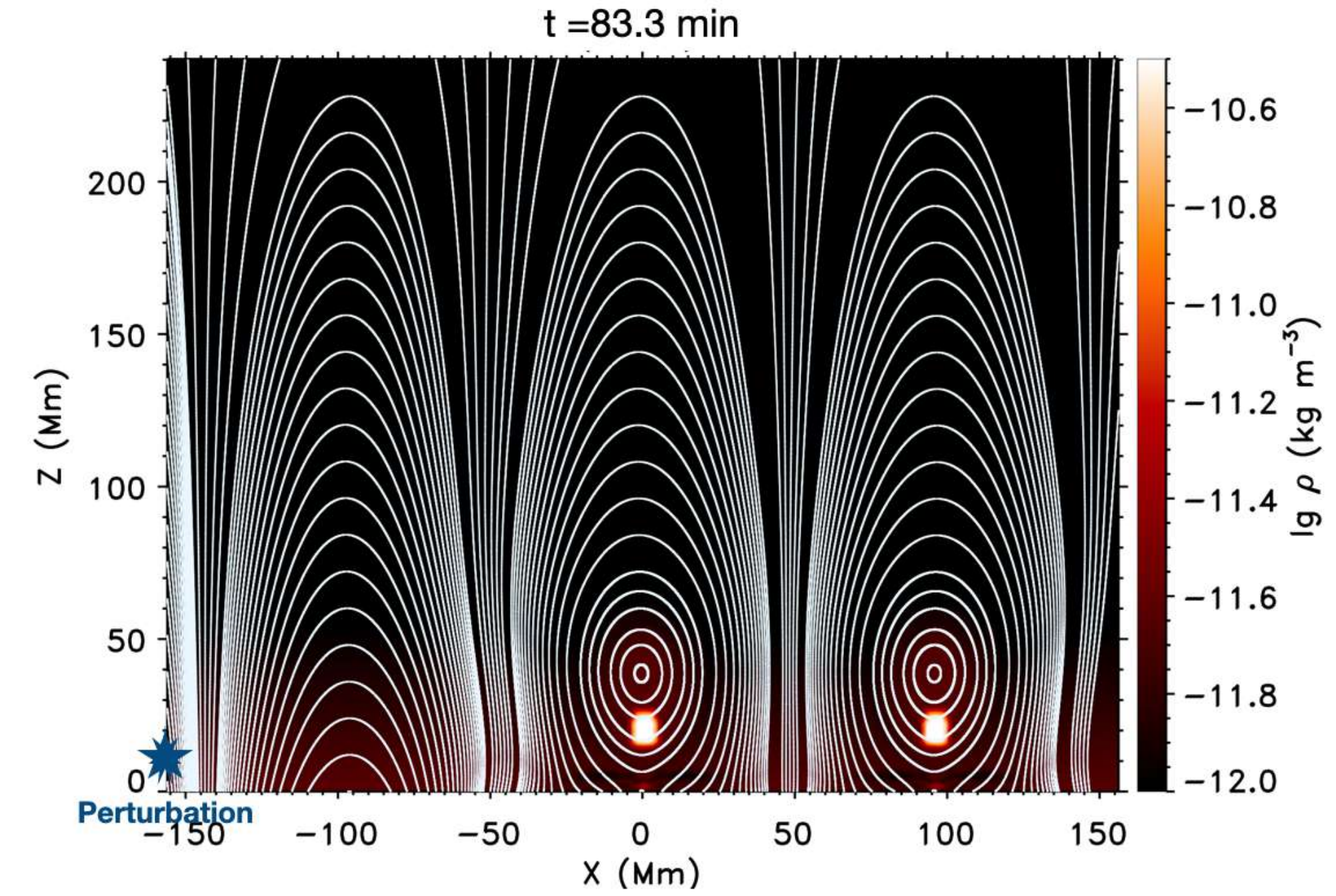}
	\caption{Density distribution at the central part of the numerical domain immediately before applying the external disturbance. The white lines show the magnetic field lines. The position of the external disturbance is also shown at the bottom left. \label{fig:external_initial_flux_ropes}}
\end{figure}
The numerical experiment in this section is motivated by the interesting observation by \citet{Shen:2014apj1}.
They reported oscillations in different filaments excited sequentially by a perturbation produced in a distant flare.
Therefore, we aim to understand how a single perturbation propagates in the magnetic environment reaching two filaments at different distances from the source of the energetic disturbance. 
Additionally, we aim to study how different polarizations of oscillations are excited by this perturbation in each prominence.

The initial configuration is the same as shown in Fig. \ref{fig:eruption-initial}, but now we form two flux ropes at $x=0$ and $x=96\Mm$ (Fig. \ref{fig:external_initial_flux_ropes}). These ropes are formed with converging motions at their respective base. The converging velocities are defined by Eq. (\ref{imposed_velocity_vx_1}). The flux ropes are formed up to time $4000\secs$, and the prominences are loaded and relaxed up to time $t=5000\secs$. 
Instead of producing an eruptive event to perturb the prominences, we used a source term in the energy equation, similar to \citet{Liakh:2020aap}. The reason has been the difficulty of producing energetic enough perturbations in Sec. \ref{sec:eruption}. We mimic the effect of a flare or energetic burst by placing a source term in the energy equation defined as
\begin{equation}\label{external_energy_source}
	S_{e}=\frac{\alpha}{t_{\rm pert}}\exp\left({-\frac{(x-x_{\rm pert})^2}{\sigma_x^2}-\frac{(z-z_{\rm pert})^2}{\sigma_z^2}}\right)\, ,
\end{equation}
where the parameter $\alpha=2\ \mathrm{J\ m^{-3}}$, and the rest of parameters $t_{\rm pert}=50\secs$, $x_{\rm pert}=-151.2 \Mm$, $z_{\rm pert}=12 \Mm$, $\sigma_{x}=\sigma_{z}=12  \Mm$ correspond to the half-size and the $x$- and $z$-coordinates of the center of the perturbation, respectively.
The positions of the prominence-hosting flux ropes and the location of the perturbation are shown in Fig. \ref{fig:external_initial_flux_ropes}. The numerical domain extends from $x=-288$ to $x=288\Mm$. The artificial perturbation approach allows us to choose the perturbation position at significant distances from the flux ropes to mimic a distant flare (see Fig. \ref{fig:external_initial_flux_ropes}). Alongside, the source of the disturbance is located far enough from the boundaries. As we use a periodic system, the undesirable perturbation appears on the other side of the domain when the perturbation reaches the left boundary. Thus, by choosing the current position, the disturbance is located at the optimal position with respect to the flux ropes and the left boundary.

\subsection{Propagation of the perturbation}

 The propagation of the non-linear disturbance in the arcade system and through the flux ropes is very complex. In order to identify the different waves propagating in the medium, we use a decomposition proposed by \citet{Cally:2017mnras},  \citet{Khomenko:2018aap}, and \citet{Khomenko:2019apj}, and  it is based on the construction of the following quantities:
\begin{eqnarray}
	f_{\rm long}&=&\mathbf{\hat{e}_{\parallel}} \cdot \mathbf{\nabla} (\mathbf{v} \cdot \mathbf{\hat{e}_{\parallel}})\, , \label{eq:f_long}\\
	f_{\rm fast}&=&\mathbf{\nabla}\cdot(\mathbf{v}-\mathbf{\hat{e}_{\parallel}}v_{\parallel}) \, ,  \label{eq:f_fast}
\end{eqnarray}
where, $\mathbf{\hat{e}_{\parallel}}=\mathbf{B}/B$ is a unit-vector directed along the magnetic field, $v_{\parallel}$ is the longitudinal velocity defined by Eq.  (\ref{long-velocity}). Since we consider low-$\beta$ plasma, the decomposition allows us to distinguish the slow and fast mode waves due to anisotropy in the propagation directions.

\begin{figure*}[!ht]
	\centering
	\includegraphics[width=0.97\textwidth]{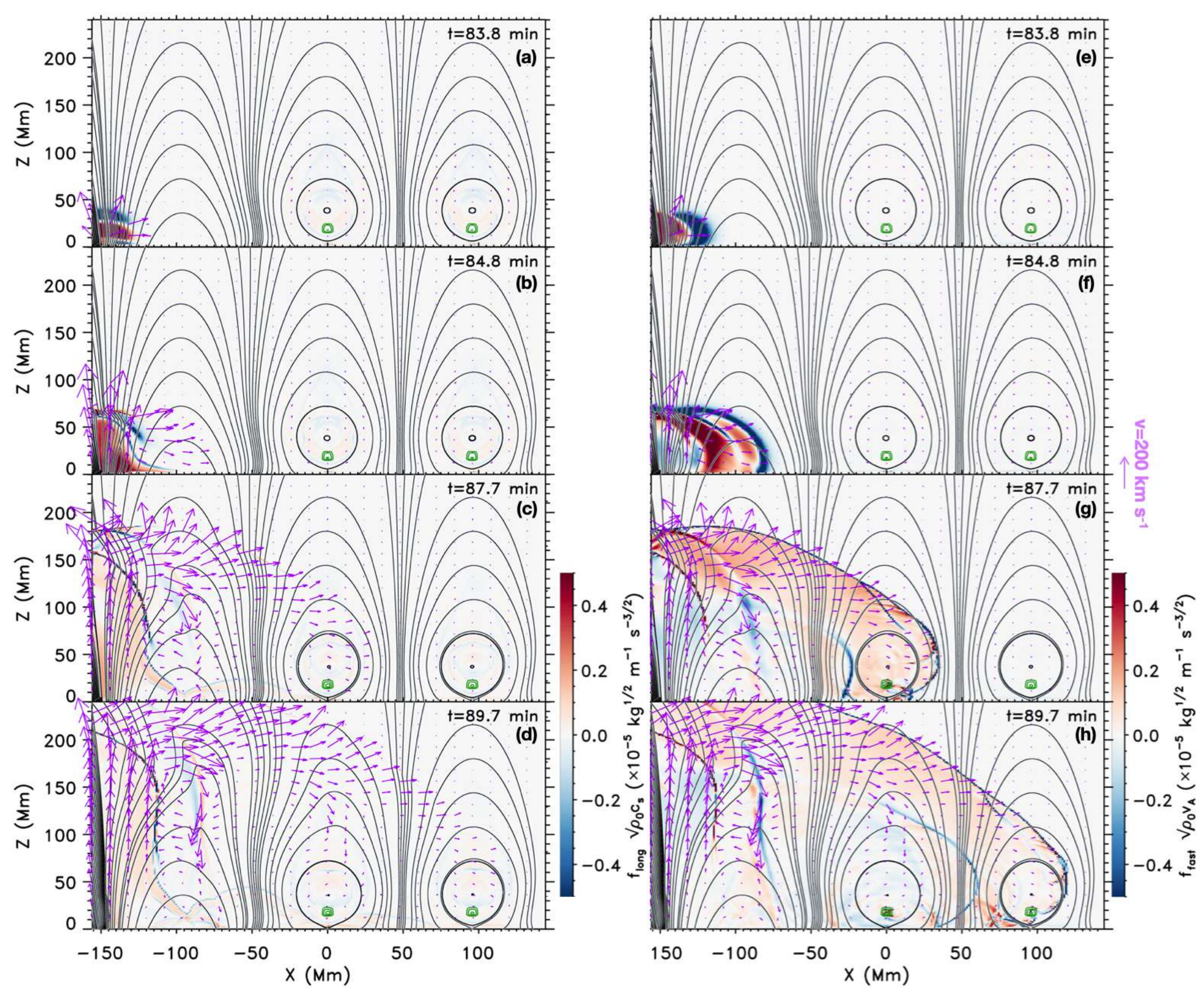}
	\caption{Temporal evolution of the wave field after initiating the perturbation (saturated). Left panels: $f_{\rm long}\sqrt{\rho_0 c_s}$; right panels: $f_{\rm fast}\sqrt{\rho_0 v_A}$. The black lines denote the magnetic field lines. The green lines show the density isocontrours. The purple arrows show the velocity field.
		\label{fig:decomposition1_flux_ropes}}
\end{figure*}

Figure \ref{fig:decomposition1_flux_ropes} show $f_{\rm long}$ scaled to $\sqrt{\rho_0 c_s}$ (left column) and $f_{\rm fast}$ scaled to $\sqrt{\rho_0 v_A}$ (right column), where $\rho_0$, $c_s$, and $v_A$ are the density, the sound and Alfv\'{e}n speed of the initial atmosphere.  Figure \ref{fig:decomposition1_flux_ropes}a shows the initiation of the disturbance. The source in the energy equation causes the disturbance that propagates in all directions (purple arrows). Figures \ref{fig:decomposition1_flux_ropes}b-d show the propagation of the slow-mode wave. After the initiation, the wave front starts to move along the vertical magnetic field. In Fig. \ref{fig:decomposition1_flux_ropes}b, the wave front reaches approximately the height of $60\Mm$. In Figs. \ref{fig:decomposition1_flux_ropes}c and d, we observe the corresponding wave front as a very thin structure at the height of $160\Mm$ and $210\Mm$, respectively. The comparison between these panels allows us to estimate the propagation speed of this front, which comes to be around $575\kms$. This value significantly exceeds the value of the local sound speed of $180\kms$. Therefore, this disturbance is significantly supersonic, and we can observe the formation of the shock front in Figs. \ref{fig:decomposition1_flux_ropes}c and d.  

Figures \ref{fig:decomposition1_flux_ropes}e and f show the initial propagation of the fast-mode wave from the perturbation region. We can observe how the wave front moves across the magnetic field of the nearby arcades in   \ref{fig:decomposition1_flux_ropes}f. Figure \ref{fig:decomposition1_flux_ropes}g shows when the fast-mode wave passes the flux rope at $x=0\Mm$. This closer flux rope appears to be pushed down and slightly inclined to the right by the front. The front affects the magnetic field considerably. Due to its propagation, the vertical magnetic field lines at $x=-48\Mm$ and the magnetic arcades around the flux rope at $x=0\Mm$ are strongly deformed. At the same time, we do not observe any significant displacement of the prominences mass. Therefore, the wave front does not appear to move the prominence mass directly. However, it significantly affects the hosting magnetic field. Considering that the plasma is mostly frozen in the magnetic field, we can expect the prominence motions as the secondary effect of the perturbation. 
Recently \citet{Liakh:2020aap} showed that the perturbation of the magnetic field of the flux rope could lead to the displacement and following excitation of the oscillations of the prominence plasma in the transverse and longitudinal directions with respect to the magnetic field. Moreover, \citet{Liakh:2021aap} found that the energy and momentum exchange between the field lines can amplify the oscillations in some field lines with the simultaneous attenuation of the oscillations in others. Similarly, in our current experiment, the magnetic flux ropes are first affected by the wave front. Then, the magnetic field gradually transfers the energy to plasma, and oscillations are induced.

In Fig. \ref{fig:decomposition1_flux_ropes}h, one can observe that the front reaches the second flux rope at $x=96\Mm$. The wave front produces a  disturbance that pushes down and inclines this flux rope, similar to the first one. Comparing Figs. \ref{fig:decomposition1_flux_ropes}g and h, we see that the wave front reaches the closer and farther flux ropes at $t=87$ and $89\mins$, respectively. This time delay corresponds to the finite velocity of the perturbing wave front. From this delay, we obtained the velocity of the wave front propagation to be as large as $\sim800\kms$. From Figs. \ref{fig:decomposition1_flux_ropes}f-h, we conclude that the fast-mode shock wave causes the main perturbation to the flux ropes in this experiment.
The properties of oscillations in the closer and further flux ropes and the excitation of different modes of LAOs by the same front are discussed in Sect. \ref{sec:prominence-oscillations-flux-ropes}.

\subsection{Prominence oscillations}\label{sec:prominence-oscillations-flux-ropes}

\begin{figure*}[!ht]
	\centering
	\includegraphics[width=0.9\textwidth]{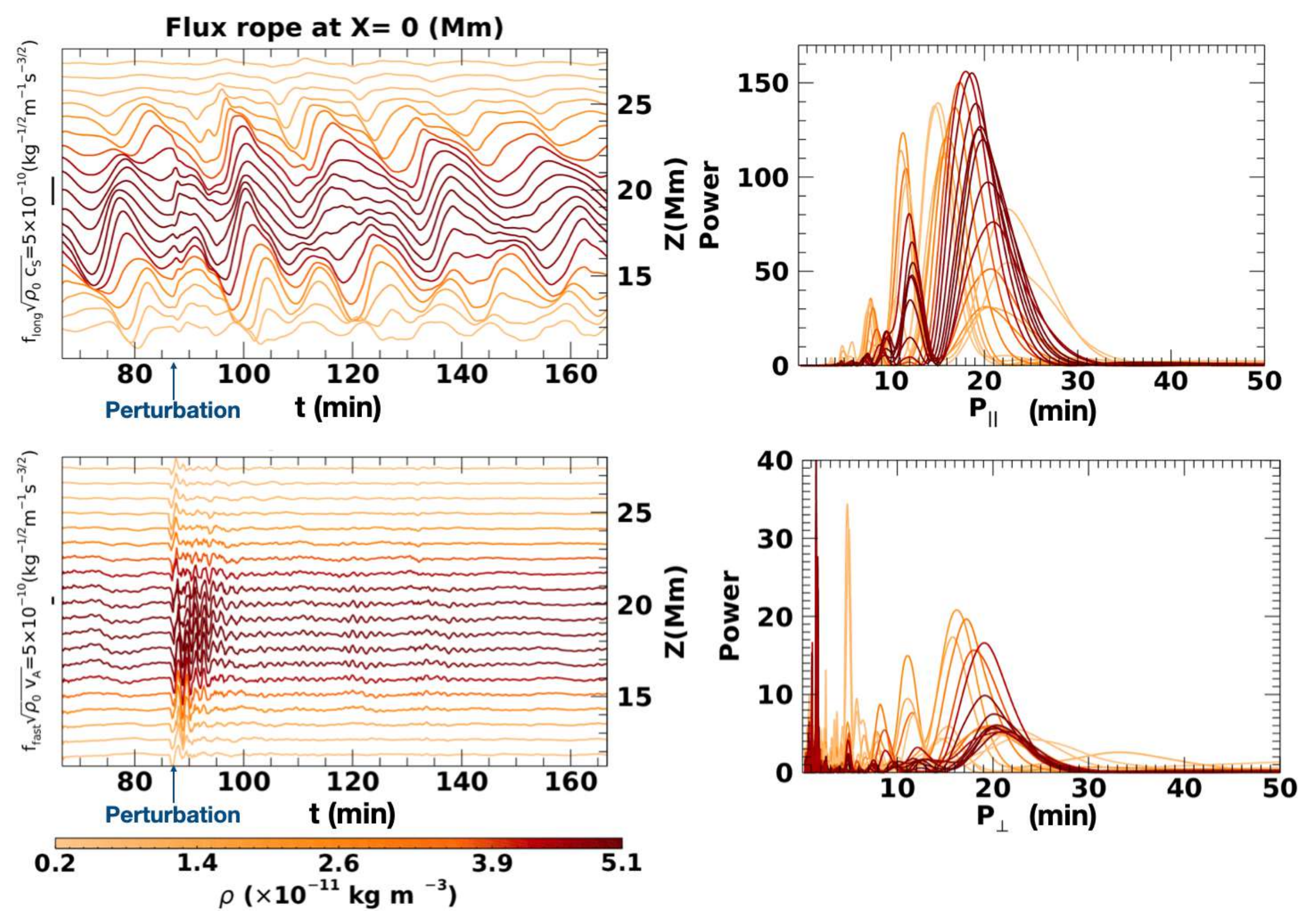}
	\caption{Temporal evolution of the wave field. Left panels: $f_{\rm long}\sqrt{\rho_0 c_s}$ (top); $f_{\rm fast}\sqrt{\rho_0 v_A}$ (bottom) at the center of mass at the selected field lines of the flux rope at $x=0\Mm$. Right panels: the periodograms of the $f_{\rm long}\sqrt{\rho_0 c_s}$ (top) and $f_{\rm fast}\sqrt{\rho_0 v_A}$ (bottom) in time-interval $83$-$167\mins$. The color bar denotes the maximum initial density at each field line. The left vertical axis indicates the velocity amplitude scale. The right vertical axis denotes the vertical location of the dips of the selected magnetic field lines. \label{fig:external_flux_rope_1_lines}}
\end{figure*}
\begin{figure}[!t]
	\centering
		\includegraphics[width=0.49\textwidth]{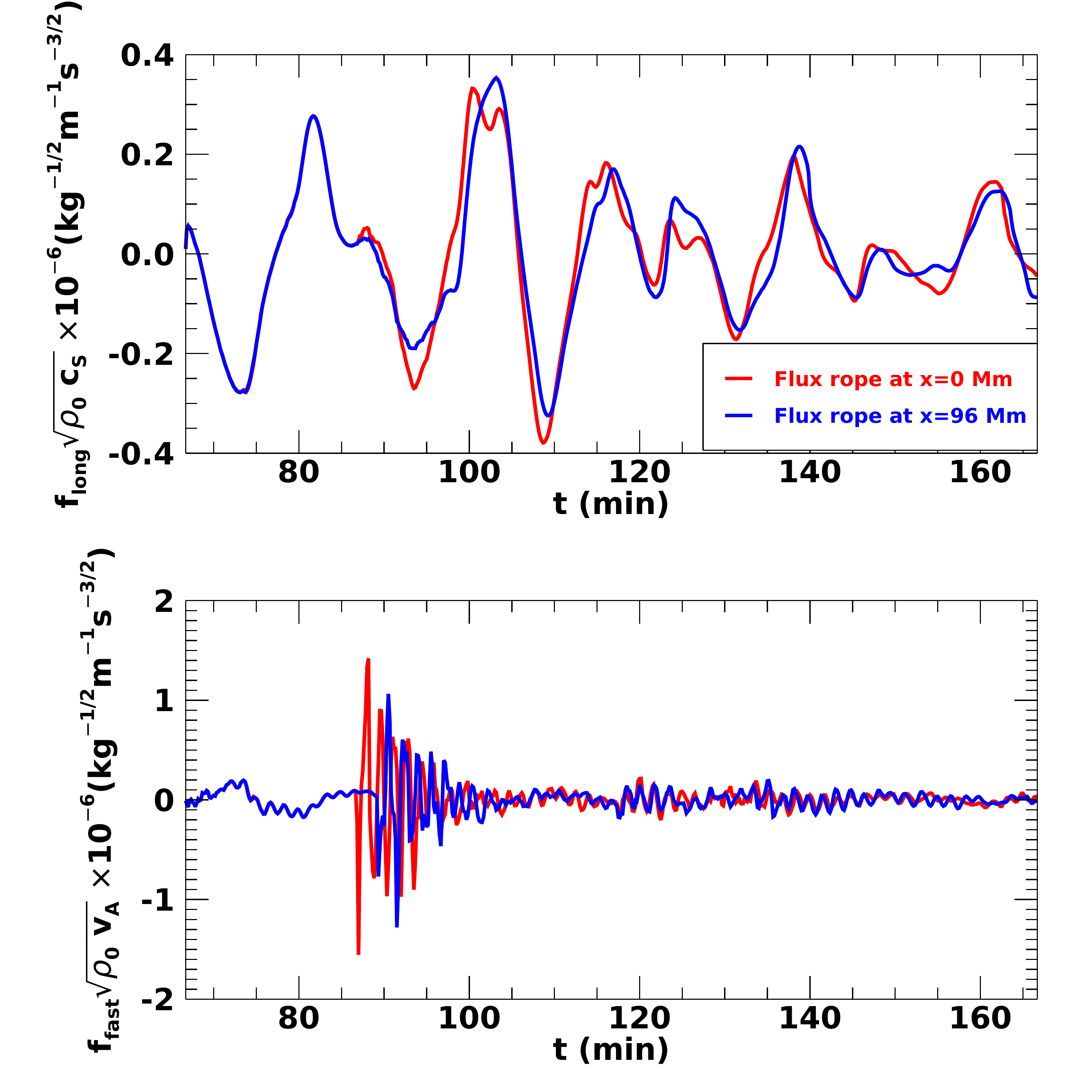}
	\caption{Temporal evolution of $f_{\rm long}\sqrt{\rho_0 c_s}$ (top) and $f_{\rm fast}\sqrt{\rho_0 v_A}$ (bottom) at the center of mass of the prominences at $x=0\Mm$  and $x=96\Mm$. \label{fig:cm_comparison}}
\end{figure}
Since the prominences magnetic field remains relatively stable in this experiment, we analyze the quantities defined by Eqs. \eqref{eq:f_long} and \eqref{eq:f_fast} at the center of mass of the selected field lines. For this analysis, we used some 20 field lines that belong to each flux rope. First, we compute the corresponding variables in the grid points of the domain. Second, we express these projected variables in the coordinates of the field lines obtained from the integration. Additionally, we interpolate the density along the same field lines. From the interpolation, we obtain the arrays for the $f_{\rm long}\sqrt{\rho_0c_S}$ and $f_{\rm fast}\sqrt{\rho_0v_A}$ and density as functions of time and position along the field lines. Finally, we compute $f_{\rm long}\sqrt{\rho_0c_S}$ and $f_{\rm fast}\sqrt{\rho_0v_A}$ at the center-of-mass of each selected field line.
First, we study the motions associated with prominence at $x=0\Mm$. Figure \ref{fig:external_flux_rope_1_lines} (left panels) shows the temporal evolution of the wave field as a function of the height of the magnetic dip, while the periodograms for the time interval $83$-$167\mins$ for these velocities are given at the right-hand side panels. The perturbation reaches the prominence located at $x=0\Mm$ at $t\sim87\mins$. The arrival of the perturbation results in the signal disturbance seen for both $f_{\rm long}\sqrt{\rho_0c_S}$ and $f_{\rm fast}\sqrt{\rho_0v_A}$. 

The top left panel shows some variation of $f_{\rm long}\sqrt{\rho_0c_S}$ before the shock wave arrival. These motions are produced as a response to the prominence mass loading to the flux rope. In the central prominence region, these oscillations demonstrate damping behavior. At $t=87\mins$, the wave perturbation disrupts the pre-existing pattern of motions. After the action of the perturbation, $f_{\rm long}\sqrt{\rho_0c_S}$ shows an amplification in the time interval $87-105\mins$. After this initial increase, the wave front significantly affects the magnetic field of the flux ropes, inclining and deforming it. As the plasma is mostly frozen in the magnetic field, the prominence mass follows the motions of the magnetic field. This way, the magnetic field transfers the perturbation to the prominence. That is why there is a delay between the action of the main disturbance and the start of the longitudinal oscillations. We applied the same analysis at the center of mass of the selected field lines to $v_{\parallel}$ and found that the former perturbation produces oscillations with an amplitude that does not exceed $5\kms$, implying the triggering of SAOs.

The periodogram of $f_{\rm long}\sqrt{\rho_0c_S}$ (top right panel of Fig. \ref{fig:external_flux_rope_1_lines}) demonstrates that the period changes from $16.0\mins$ at the top to $20.6\mins$ at the bottom of the prominence. The period of the longitudinal oscillations in this experiment and its variation with height resembles the one from the previous experiment shown in Fig. \ref{fig:eruption_vpar_vper_corks}. The reason for that can be using the same flux rope prominence model in both numerical experiments. Assuming the pendulum model \citep{Luna:2012apjl} suggests that the main restoring force for the longitudinal oscillations is the gravity projected along the magnetic field, and the period depends only on the radius of curvature of the magnetic field lines. Since we use identical prominence magnetic field structures, we obtain similar periods in both cases. To confirm that suggestion, we compute the time-average value of the radius of curvature, $R_c$, of the selected field lines. Then, using the following formulae
\begin{equation}\label{eq:period-pendulum}
	P=2\pi\sqrt{R_c/g}\, ,    
\end{equation}
where $g$ is the solar gravitational acceleration, we obtained periods $16.4-19.9\mins$ in the densest prominence region. These theoretical periods agree with those we obtained in both numerical experiments. In a flux rope, the radius of curvature decreases from the edges to the center, explaining why in our experiments, the period of longitudinal oscillation decreases with increasing height (See Figs. \ref{fig:eruption_vpar_vper_corks} and \ref{fig:external_flux_rope_1_lines}). 

Another peak in the periodogram corresponds to periods $11-12\mins$. This periodicity is less relevant for the central prominence region. Its power increases for the plasma that belongs to the corona and prominence-corona transition region. These oscillations can be related to the pressure-driven mode. As shown by \citet{Luna:2012apj2} and \citet{Liakh:2021aap} in the region with reduced density contrast and shallow lines, the gas pressure force dominates over the gravity force.

The bottom panels of Fig. \ref{fig:external_flux_rope_1_lines} show the analysis of $f_{\rm fast}\sqrt{\rho_0v_A}$ in the same prominence. When the wave front approaches the prominence ($t=87\mins$), it causes the perturbation of $f_{\rm fast}\sqrt{\rho_0v_A}$ in all the lines. Owing to this perturbation, oscillations are established. The motions at all the field lines are similar with no significant shift, as in the case of  $f_{\rm long}\sqrt{\rho_0c_S}$. These vertical oscillations show significant attenuation in the time interval $87-100\mins$. In the next minutes, we again see the signature of these oscillations. These wave trains appear several more times in the signal. \citet{Zhang:2019apj} and \citet{Liakh:2020aap} obtained similar wave trains. We concluded that these wave trains were caused by non-linear oscillatory behavior and mode coupling. Additionally, we analyze $v_{\perp}$ at the center of mass of the magnetic field lines in the experiment described in this paper and find that the perturbation shown in the bottom left panel of Fig. \ref{fig:external_flux_rope_1_lines} corresponds to the velocity amplitude around $10\kms$, implying triggering of LAOs.

The periodogram on the bottom right panel of Fig. \ref{fig:external_flux_rope_1_lines} shows that the dominant peak for $f_{\rm fast}\sqrt{\rho_0v_A}$ is located at the periodicity around $1.1-1.7\mins$. In the previous experiment with eruption, yet another period of $2$ minutes was prominent in the periodogram (Sect \ref{sec:prominence-oscillations-eruption}). Following \citet{Nakariakov:2005lrsp}, we assume these oscillations are standing kink modes in the prominence threads. Then, we compute the period using an expression:
\begin{equation}\label{eq:period-nakariakov}
	P=\frac{L}{B_0}\sqrt{2\mu_0\rho_0(1+\rho_p/\rho_0)}\, ,    
\end{equation}
where $L$ is the length, $B_0$ is the magnetic field strength prior to the perturbation, and $\rho_p$ is the prominence density of the corresponding thread. This expression also includes the background coronal density $\rho_0$. We obtained the period in the $0.9-2.0\mins$, which agrees with the periods obtained in the numerical experiments. Therefore, the short-period vertical oscillations are defined by the properties of the corresponding threads, such as the length of the thread, the magnetic field strength in the corresponding field lines, and the density contrast.

The second important peak in the periodogram is the one around $16.2-19.0\mins$. This peak is caused by the pendulum longitudinal oscillations due to mode coupling. Finally, the third important peak in the densest prominence region is at $\sim5\mins$. This periodicity resembles another periodicity at $4.4-4.6\mins$ in the eruption experiment ( bottom right panel of Fig. \ref{fig:eruption_vpar_vper_corks}). We try associating this mode with the simple harmonic oscillator model with the magnetic tension as the main restoring force \citep{Hyder:1966zap}. The authors computed the magnetic field strength from the period of the global vertical oscillations. In our case, we compute the average value of the magnetic field strength and density in the prominence, $\langle B_{0}\rangle =7.7$ G and $\langle \rho_{p}\rangle =1.87\times10^{-11}\mathrm{kg\ m^{-3}}$ and prominence height scale, $h_0=12.72$ Mm. Then, using the expression
\begin{equation}\label{eq:period-hyder}
	P=\frac{2\pi h_0}{\langle B_{0}\rangle }\sqrt{\pi\langle \rho_{p}\rangle }\, ,    
\end{equation}
we obtain the period of the global vertical oscillations $4.2\mins$, which can be related to $4$-minutes oscillations obtained in the periodograms.

We compare oscillations at the center of mass of both prominences in Fig. \ref{fig:cm_comparison}. We obtain a larger amplitude for the transverse motions (bottom panel) than for the longitudinal motions (top panel) for both flux ropes. The bottom panel shows that the perturbation affects the farther flux rope slightly later, which is related to the finite speed of the perturbing wave. The amplitude of $f_{\rm fast}\sqrt{\rho_0v_A}$ seems slightly larger for the closer flux rope, showing that the perturbing wave slightly loses energy when reaching a farther flux rope. We also find that both prominences have similar periods of longitudinal and transverse motions. For the longitudinal oscillations, we obtain a period of $\sim 20\mins$, and the vertical oscillations have a short period of $\sim 2\mins$.
This experiment confirms the excitation of a similar kind of oscillation at the flux rope prominences located at different distances from a distant perturbation. We obtained nearly the same interaction of the energetic wave with the prominences located at the distances, $151.2$ Mm and $247.2$ Mm. Therefore, our experiment can explain the triggering of the LAOs in the chain of the filaments in the observations by \citet{Shen:2014apj1}. However, it remains unclear how several filaments located close to the flare regions did not show any response in the former observational event. Such behavior can be associated with the different mutual placement of the filaments with respect to the arriving wave front. Studying the triggering of prominence motions in the 3D configuration may shed light on this aspect. This can be addressed to the future study.

\section{External perturbation in association with distant dipped arcade prominence}\label{sec:dipped-arcade}
The third experiment is motivated by the study of \citet{Shen:2014apj2}. The authors
suggested that the longitudinal oscillations of the prominence located in the dipped sheared arcade could be excited by a shock wave when the propagation direction is parallel to the filament spine. This shock wave is a coronal counterpart of the Moreton wave, a fast magnetoacoustic shock wave. In contrast with the impact of the jet, the perturbation is not associated with plasma flows. \citet{luna_large-amplitude_2021} studied the excitation of LALOs in the dipped arcade when the source of the perturbation was an energetic jet with a magnetic connection with the main prominence field. The authors found that the LALOs are excited by the direct influence of the mass flow injected by the jet. In this study, we aim to investigate the possibility of triggering LALOs when the source of the perturbation is located far enough and is not magnetically connected to the prominence magnetic field.
\begin{figure}[!ht]
	\centering
	\includegraphics[width=0.5\textwidth]{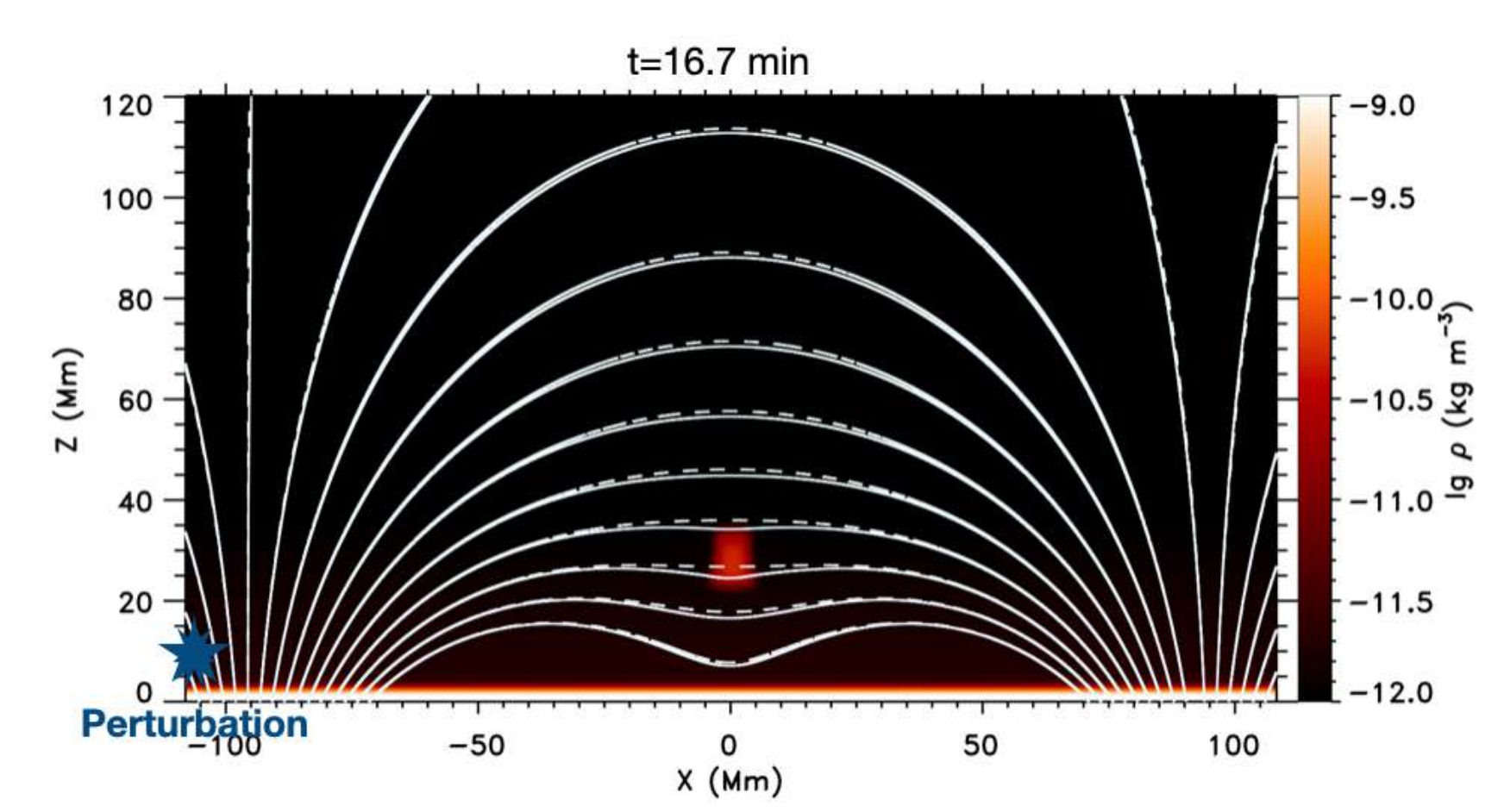}
	\caption{Density distribution and magnetic field lines at the central part of the computational domain after the mass loading. The dashed lines denote the initial magnetic field before the mass loading. The position of the external disturbance is also shown at the bottom left. \label{fig:external_double_initial}}
\end{figure}
\begin{figure*}[!ht]
	\centering
	\includegraphics[width=0.97\textwidth]{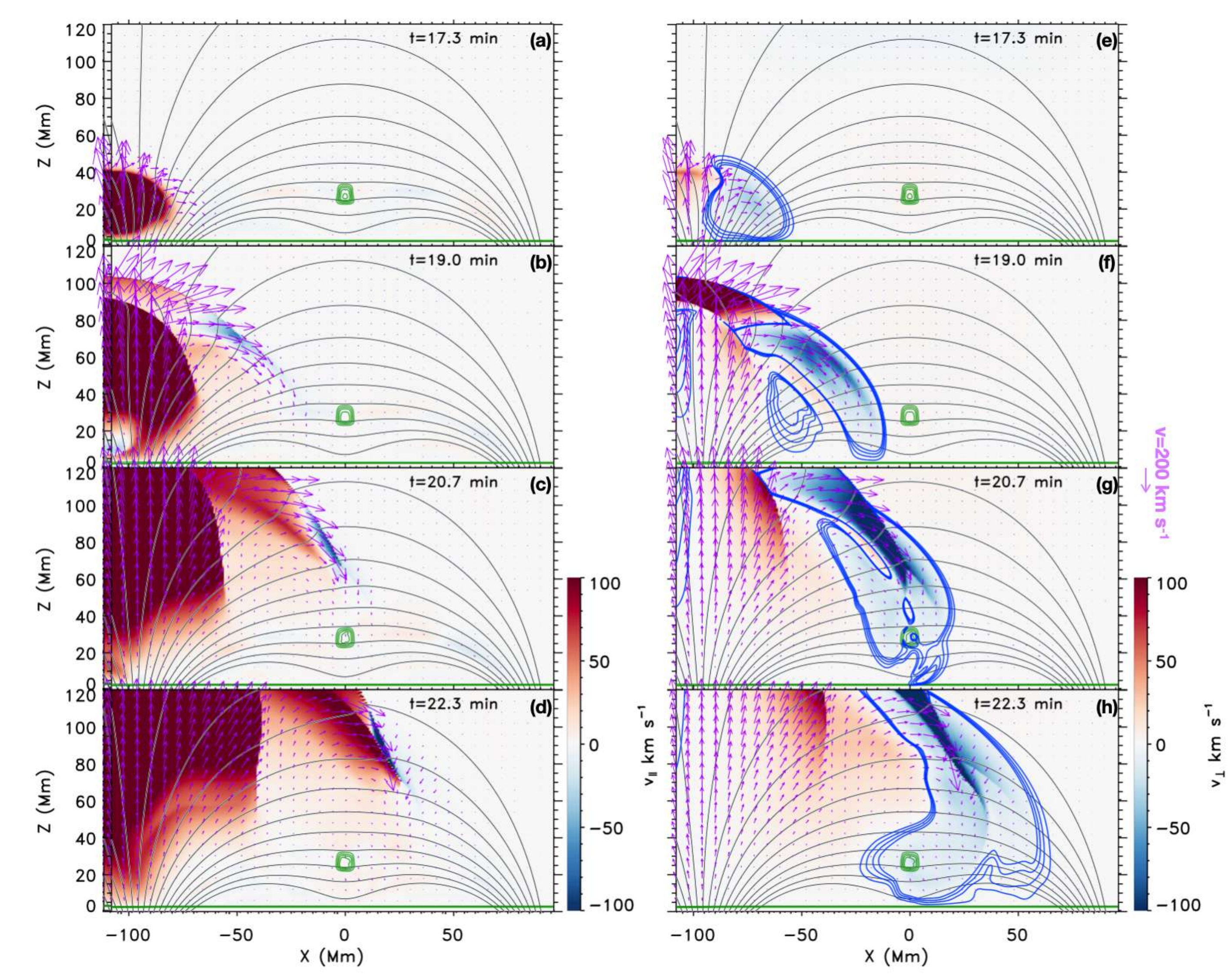}
	\caption{Evolution of $v_{\parallel}$ (left panel) and $v_{\perp}$ (right panel) during the wave propagation (saturated). The black lines denote the magnetic field. The green lines show the density isocontours. The blue lines show the $v_{\perp}$ isocountours. The purple arrows denote the velocity field.\label{fig:external_double}}
\end{figure*}
\begin{figure*}[!ht]
	\centering
	\includegraphics[width=0.97\textwidth]{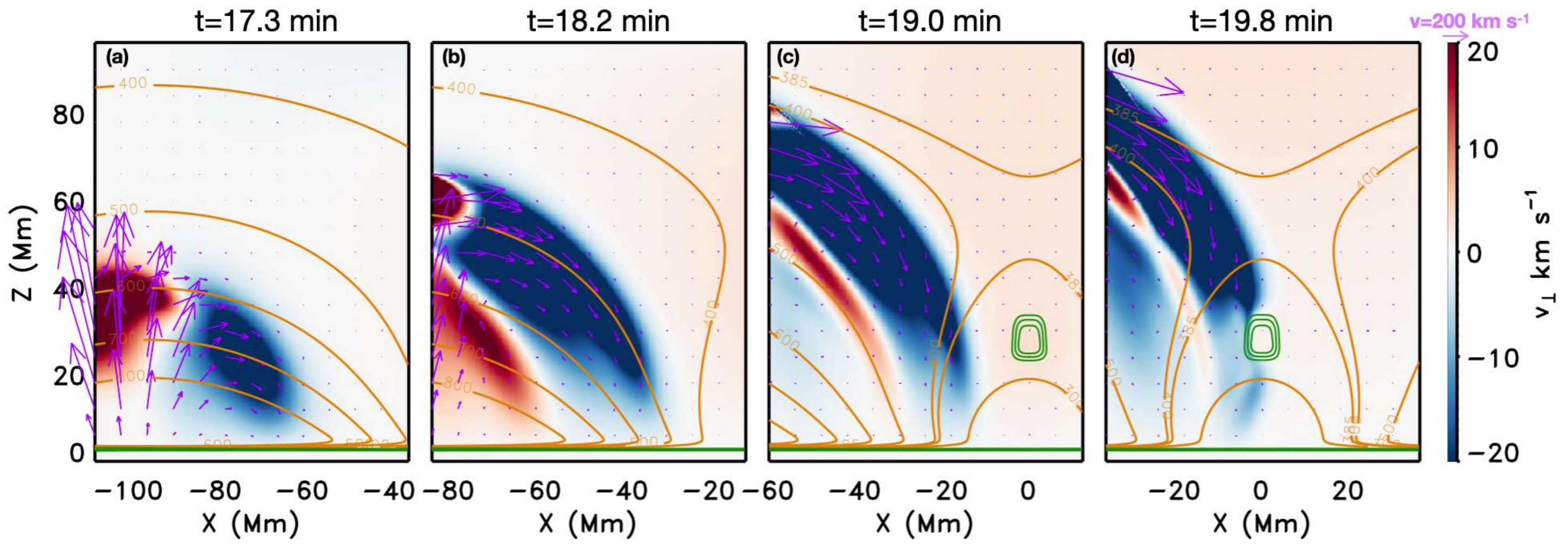}
	\caption{Evolution $v_{\perp}$ during the wave propagation (saturated). The orange lines denote the isocontours of the phase speed $v_{ph}=\sqrt{v_A^2+c_S^2}$. The green lines show the density isocontours. The purple arrows denote the velocity field.\label{fig:external_double_front}}
\end{figure*}
\begin{figure*}[!ht]
	\centering
	\includegraphics[width=0.9\textwidth]{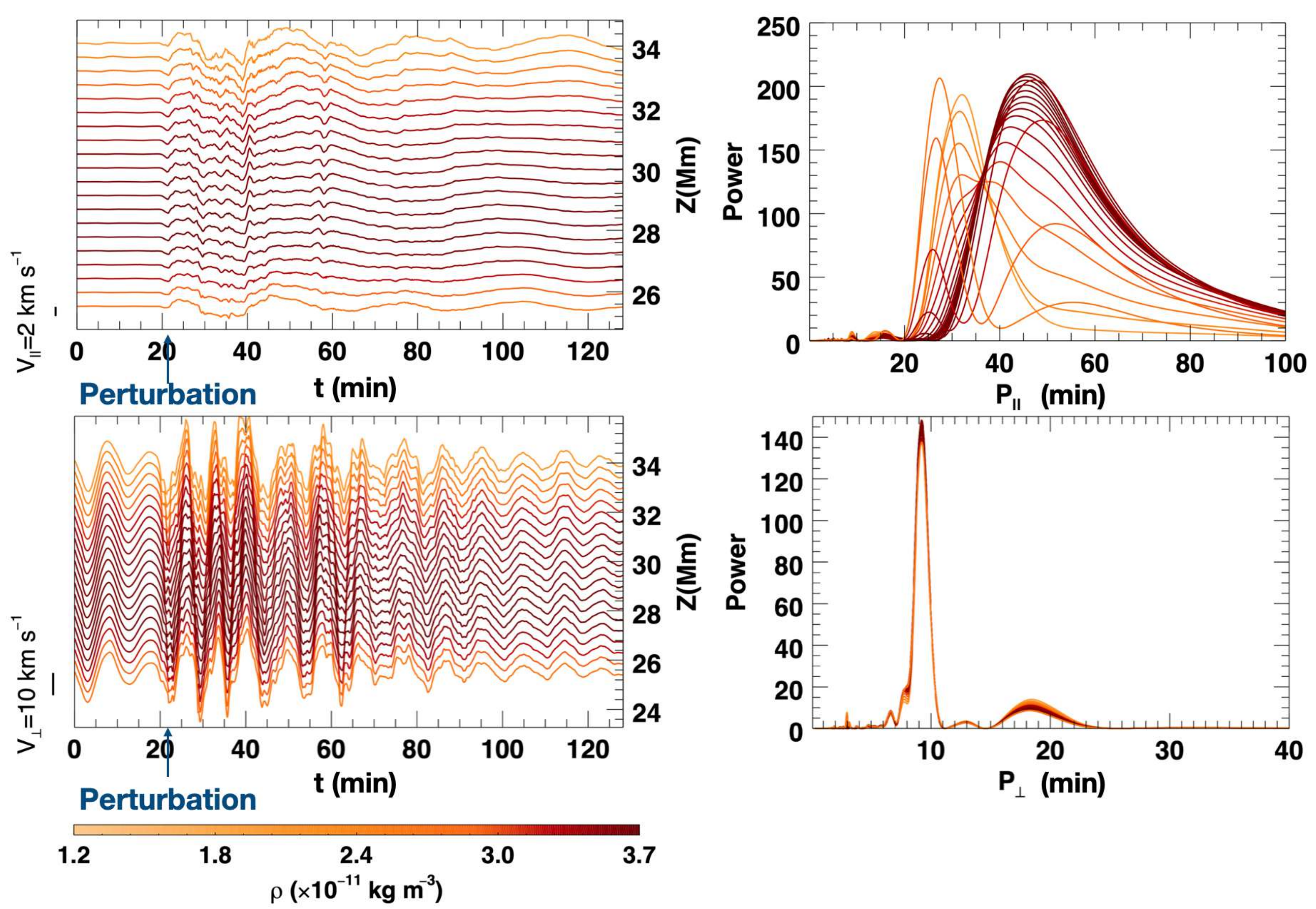}
	\caption{Temporal evolution of the velocity field in the prominence region. Left panels: $v_{\parallel}$ (top) and $v_{\perp}$ (bottom) at the center of mass at the selected field lines of the dipped arcade. Right panels: the periodograms of the $v_{\parallel}$ (top) and $v_{\perp}$ (bottom) in time-interval $22$-$127\mins$. The color bar denotes the maximum initial density at each field line. The left vertical axis indicates the velocity amplitude scale. The right vertical axis denotes the vertical location of the dips of the selected magnetic field lines.\label{fig:external_double_lines}}
\end{figure*}

The numerical 2D setup, shown in Fig. \ref{fig:external_double_initial}, is similar to our previous study \citet{Liakh:2021aap}. The numerical domain consists of a box of $384 \times 240\Mm$ in size (spatial resolution $240$ km). The initial magnetic field is potential and consists of a superposition of two arcades forming a dipped part at its center
\begin{eqnarray}\label{magnetic-components-dipped}
	B_{x}(x,z)/B_0&=&\cos k_{1}x \, e^{-k_{1}(z-z_{0})}-\cos k_{2}x \, e^{-k_{2}(z-z_{0})} \, , \\
	\label{magnetic-component-z}
	B_{z}(x,z)/B_0&=&-\sin k_{1}x \, e^{-k_{1}(z-z_{0})}+ \sin k_{2}x \, e^{-k_{2}(z-z_{0})},
\end{eqnarray}
where we have considered the parameters $z_{0}=-2\Mm$, $B_{0}=10$ G,  $k_{1}=\pi/D$ and $k_{2}=3k_{1}$ where $D=191.4\Mm$ is the spatial periodicity of the magnetic structure along the $x$-direction. The magnetic field strength varies between $3$ and $4$ G from the bottom to the top of the dipped region.
Similarly to our previous work, the initial atmosphere is a stratified plasma in hydrostatic equilibrium, including the upper chromosphere, transition region (TR), and corona. The corresponding temperature profile is written as
\begin{equation}\label{eq:temperature-profile_external}
	T(z)= T_{0} + \frac{1}{2} \left(T_\mathrm{c}-T_{0} \right) \left[1+ \tanh
	\left( \frac{z-z_{c}}{W_z}\right) \right] \, . 
\end{equation}
We choose $T_\mathrm{c}=10^{6}$ K, $T_{0}=10^{4}$ K, $W_z=0.7\Mm$, and $z_{c}=4.8\Mm$. This profile provides the temperature ranging from $T_ {ch}=10^{4}$ K at the base of the chromosphere to $T_{c}=10^{6}$ K in the corona. As the plasma is stratified in the vertical direction, the density changes from $\rho=9\times 10^{-9}\ \mathrm{kg\ m^{-3} }$ in the chromosphere to $\rho=1.98\times 10^{-12}\ \mathrm{kg\ m^{-3} }$ at the base of the corona at the height $z_{c}=4.8\Mm$. In contrast to the two previous experiments, the initial atmosphere in this numerical setup includes lower layers of the solar atmosphere, i.e., the chromosphere and TR. On the one hand, this helps to satisfy the line-tying condition of the magnetic field. On the other hand, this allows us to reproduce a more realistic propagation of the shock wave in the different layers of the solar atmosphere, implying the inclination of the wavefront due to the variation of the local phase speed with height \citep{Uchida:1968solphys, Liu:2012apj, RiuLiu:2013apj}.

We impose the periodic condition at the side boundaries and the current-free condition for the magnetic field and zero velocities as in \citet{Terradas:2013apj}, together with the symmetric condition for the temperature and pressure and fixed density at the bottom. At the top boundary, we use the symmetric boundary condition for all the variables except for $B_{x}$, which is antisymmetric.

We load the prominence mass at the dipped region of the magnetic field by increasing the density using a source term in the continuity equation (Eq. \eqref{eruption_mass_source}) defined by a Gaussian distribution centered at $(x,z)=(0,30)\Mm$. The mass loading starts at $t=0 \secs$ and ends at $t=100 \secs$. The prominence density is set to $30$ times larger than in corona, and the prominence dimensions are $7$ and $10\Mm$ in horizontal and vertical directions.
During the first  $16.7\mins$, we let the magnetic configuration with prominence evolve, reaching an equilibrium state. In Fig. \ref{fig:external_double_initial}, the solid lines show the magnetic field after the relaxation process, and the dashed lines illustrate the initial magnetic field. As we can see, the magnetic field is slightly deformed with respect to the initial configuration due to the heavy mass. 

The external perturbation is produced by the source term in the energy equation, similar to the previous section, with  $\alpha=2\ \mathrm{J\ m^{-3}}$, $t_{\rm pert}=50\secs$, $x_{\rm pert}=-105.6 \Mm$, $z_{\rm pert}=12 \Mm$ and $\sigma_{x}=\sigma_{z}=12  \Mm$. The approximate position of the perturbation is shown by the blue star in Fig. \ref{fig:external_double_initial}. This position is not magnetically connected to the prominence to mimic the effect of a distant flare.

Since we are in the low-$\beta$ regime, the anisotropy in the propagation of the fast and slow modes allows us to distinguish both fronts by simply using the temporal evolution of $v_{\parallel}$ and $v_{\perp}$ instead of the less intuitive magnitudes given by Eqs. \ref{eq:f_long}-\ref{eq:f_fast}. 
Fig. \ref{fig:external_double} shows the time evolution of the longitudinal (left panels) and transverse (right panels) velocities. Figs. \ref{fig:external_double}a and e represent the time when the perturbation is applied. The perturbation starts propagating mainly vertically but also in the horizontal direction (purple arrows).
As time advances, the fast wave, $v_{\perp}$, moves ahead of the slow mode, $v_{\parallel}$. The magnetic field lines are strongly deformed on the passage of the fast wave front. In panel b, we see no relevant $v_{\parallel}$ moving toward prominence. In contrast, in panel f, some $v_{\perp}$ is directed to the cold mass. The $v_{\perp}$ isocountours (blue lines) reflects the orientation of the fast wave front. In panel c, we see that the slow mode has indeed avoided the prominence, while in panel g, part of the fast front has reached it. The velocity isocountours in panels f and g show that although the incident wave is along the prominence spine from the global view, the wave front is not strictly oriented along the spine in the moment of interaction. Moreover, when the perturbation reaches prominence,  the wave front appears oblique. This detail can play an important role in the type of induced oscillations. According to the density isocontours, the prominence body is slightly deformed due to wave propagation. In panels d and h, it can be seen that both disturbances move toward the other end of the system, moving mainly through the overlying arcade.

Figure \ref{fig:external_double_front} gives a closer look at the fast wave front and their interaction with the prominence. After the initiation of the perturbation, the front propagates in an environment with an important variation of the phase speed of the fast magnetoacoustic mode. 
Fig. \ref{fig:external_double_front}a shows the perturbation right after the energy release. Note that the phase speed is computed with unperturbed values. The perturbation moves in a medium with decreasing phase speed with height.
As it moves away and through the magnetic structure, the front becomes vertical, and the velocity field (purple arrows) turns into vertical, pointing downwards (Figs. \ref{fig:external_double_front}b-d).
In panels c and d, the bottom part of the front reaches the region where the phase speed is in the range of $300-400\kms$ while the upper part is still in the region with the characteristic speed $400-500\kms$. In this way, the lower front moves slower in contrast to the upper part, and the inclination of the front change, becoming more vertical. Panel d shows the moment when the front interacts with the prominence. The perturbation is directed mainly downwards, thus, the prominence is pushed downwards, and the oscillations are initiated. 

We analyze both longitudinal and transverse motions at different field lines as in \citet{Liakh:2021aap}. We consider a set of field lines. These lines have the initial position located at the bottom boundary. Then, we integrate the field line each time moment. As the integration is from positions with zero velocity, any field line follows the plasma motion, simplifying the study of the oscillations. For every line, we compute the density-averaged $v_{\parallel}$ and $v_{\perp}$. The results of the analysis of $v_{\parallel}$ are shown at the top panels of Fig. \ref{fig:external_double_lines}. 
As discussed above, the disturbance acts mostly as a downward perturbation. From the top left panel, we see that the velocity amplitude of the longitudinal component is very small. We can only detect oscillations with a small amplitude below $2\kms$, thus, no LALOs in this prominence are excited.

The left bottom panel of Fig. \ref{fig:external_double_lines} shows the density-averaged $v_{\perp}$. As the field is horizontal in the central part of the prominence, this movement corresponds to a vertical oscillation. We see that before the arrival of the disturbance, there were already vertical oscillations. These are associated with the mass loading process and are difficult to eliminate because they have a very long damping time. When the disturbance arrives at $t=21.7\mins$, it can be clearly distinguished that the induced oscillations have a larger amplitude than the pre-existing motions. These oscillations are synchronized at the different field lines, as seen in the panel. 
The bottom right panel of Fig. \ref{fig:external_double_lines} shows their periodograms. The dominant period at all the lines is approximately equal to $9\mins$. The vertical period remains constant with height, indicating a global normal mode. As the prominence shows the global vertical motions, it is relevant to compare the $9$-minutes period with the period given by the \citet{Hyder:1966zap} model using Eq. (\ref{eq:period-hyder}). Using averaged values of the magnetic field strength $\langle B_{0}\rangle =4$ G, averaged prominence density $\langle \rho_{p}\rangle =1.99\times10^{-11}\mathrm{kg\ m^{-3}}$, and its scale height $h_0=13.2$ Mm; we obtain a period of $P=8.6\mins$ in agreement with the value obtained in the experiment.

\section{Summary and conclusions}\label{sec:summary}

In this paper, we study the mechanisms of the triggering of LAOs in prominences by external disturbances.
We explore three scenarios representing situations that have already been observed.
This study is based on time-dependent 2.5D and 2D numerical simulations performed with the MHD code MANCHA3D. 

In the first experiment, we investigated the excitation of oscillations in the flux rope prominence by the eruption of the nearby flux rope.
Eruptive events are considered the main sources of the Moreton and EIT waves production \citep{Biesecker:2002apj,Chen:2006apj, Chen:2011spr}. These waves can propagate over large distances in the solar atmosphere and are proposed to be responsible for the excitation of LAOs \citep[see, e.g.,][]{Shen:2017apj,Devi:2022adsr}. 
However, in our experiment, we observe that the eruption itself does not excite LAOs in prominence nor find clear evidence for a  Moreton wave reaching prominence. 
Perhaps, in our configuration, the conditions for a Moreton wave to form are not fulfilled. The background magnetic field is too strong at the bottom, and the wave is refracted toward the high corona. Additionally, the prominence is very close to the reconnection site and hence is significantly influenced by the reconnection inflow. However, the configuration used in this experiment is still applicable for the study of the case where two active-region prominences are located nearby.
A strong perturbation appears at the front of the erupting flux rope in the form of density enhancement followed by dimming, moving laterally and upwards. In principle, such a front can be identified as an EIT wave \citep{Chen:2002apjl}. However, in our case, this EIT front is mainly located above the prominence and, therefore, cannot affect it. 
In order to improve this erupting model in the context of the triggering efficiency, a stronger magnetic field is needed, and a flux rope erupting lower in the corona as, for instance, in the catastrophic scenario \citet{Chen:2002apjl} and \citet{Zhao:2022apj}. 

An elongated current sheet is formed behind the erupting flux rope during the eruption. 
The reconnection inflows towards the current sheet also affect the prominence magnetic field producing an inclination of the structure. In addition, the current sheet becomes unstable, and plasmoids start to form in it.
After these plasmoids are formed, they move downwards to the post-reconnection loops or upwards following the erupting flux rope. 
Our experiment shows that the latter ones do not significantly affect the nearby filament. In contrast, the plasmoids propagating downward reach the small post-reconnection loops. The interaction of the plasmoids with these loops produces the perturbation of the velocity field. This velocity perturbation propagates in the surroundings and enters into the flux rope, causing the disturbance of the prominence mass. The analysis of the oscillatory motions of the prominence plasma in the flux rope shows that only SAOs are excited due to the nearby eruption and the plasmoid instability. The motions have a complex character showing a mixture of longitudinal and vertical oscillations with short and long periods. 
It has been known from the numerical experiments that tearing instability can develop in the current sheet. As a result, the plasmoids can be formed. 
The activity of the plasmoids is also considered a candidate for generating quasi-periodic propagating fast-mode magnetoacoustic waves (QPFs). In this scenario, the disturbances are generated by the precipitation of the plasmoids at the foot points \citep{Yokoyama:1998,Yang:2015apj,Barta:2008aap}. 
In our work, these disturbances are responsible for the excitation of the SAOs in the nearby prominence. 

In the second experiment, we partially reproduce the scenario described in \citet{Shen:2014apj1}. 
We study the effect of the same energetic disturbance on two different filaments at different distances from the source of the perturbation. The analysis reveals a fast-mode wave front propagates across both magnetic flux ropes. However, as we found in \citet{Liakh:2020aap}, the first front does not trigger the oscillation. It is the
displacement and deformation of the magnetic environment of the rope that produces the mass displacements from its equilibrium and then its oscillations. The study of oscillatory modes shows transverse oscillations with similar dominant periods in both flux ropes. The periods of the transverse oscillations in the threads agree with the model of the standing kink mode \citet{Nakariakov:2005lrsp}. Additionally, we obtained longitudinal SAOs with similar oscillatory properties in both flux ropes. In both prominences, the period of the longitudinal oscillations changes with height due to the different curvature of the magnetic field lines in good agreement with the pendulum model \citep{Luna:2012apjl}.
Similarly to our previous study \citet{Liakh:2020aap}, the induced longitudinal oscillations have a smaller amplitude than the vertical oscillations. It is very interesting that in this experiment, the perturbation produces similar oscillations in both prominences regardless of their distance from the source. This indicates that similar energy is deposited in both structures. This would show why these perturbations can produce LAOs even in filaments far away from the flare. However, it is necessary to explore more numerical cases, including three-dimensional scenarios.

Finally, in the third case, we have studied the external triggering of longitudinal LAOs in a dipped arcade model. \citet{Shen:2014apj2} suggested that a shock can excite LALOs when the shock propagates along the filament spine.
Our numerical experiment shows that a distant perturbation, not magnetically connected with prominence, can reach prominence and strongly affect the magnetic field of the arcades. When the wave front reaches prominence, it strongly pushes down the dipped region with the prominence mass inside. This way, the global vertical oscillations are established. The analysis reveals that these motions have a large amplitude that exceeds $10\kms$. The vertical period remains constant with height suggesting a global normal mode. This period is in agreement with the model of the harmonic oscillator with magnetic tension as a main restoring force \citep{Hyder:1966zap}. Additionally, the prominence evolution shows evidence of the motions along the magnetic field due to compression and rarefaction. Notably, we do not observe the excitation of LALOs from the analysis of motions. Only transverse LAOs and motions due to compression and rarefaction are excited in this numerical experiment.

This experiment shows the difficulty of directly exciting LALOs by external perturbations when propagating along their spine. This does not rule out that they can be excited but perhaps indirectly, as seen in the case of flux ropes \citep{Liakh:2020aap}. In that situation, the shock disturbs the magnetic structure that eventually moves the prominence. However, three-dimensional simulations of magnetic ropes are needed to understand this process, and this will be left for future research.

From all the experiments described above, we conclude that investigating the triggering mechanism of LAOs is crucial for understanding the physical nature of prominences and the propagation and interaction of energetic waves with different magnetic configurations. We have studied the triggering of LAOs by an eruptive event and by the artificial energy release on the flux rope and dipped magnetic arcade configurations. We conclude that the external triggering of prominence oscillations is a complex process that excites longitudinal, transverse, or a mix of both types of motions with small or large amplitudes. 

In the future, the external triggering mechanism should be investigated in 3D models. This would allow us to investigate in more detail which polarizations are excited when the perturbation comes from a different direction with respect to the axis of the prominence spine. The numerical model of the interaction of the erupting flux rope with prominence should be improved by increasing the distance between both objects or reducing the size of the shock front from the eruption. The role of plasmoids in triggering the LAOs due to nearby eruptions should be further investigated.

\begin{acknowledgements} 
	V. Liakh is supported by the ERC Advanced Grant PROMINENT from the European Research Council (ERC) under the European Union’s Horizon 2020 research and innovation programme (grant agreement No. 833251 PROMINENT ERC-ADG 2018). M. Luna acknowledges support through the Ramón y Cajal fellowship RYC2018-026129-I from the Spanish Ministry of Science and Innovation, the Spanish National Research Agency (Agencia Estatal de Investigación), the European Social Fund through Operational Program FSE 2014 of Employment, Education and Training and the Universitat de les Illes Balears. E. K. thanks the support by the European Research Council through the Consolidator Grant ERC-2017-CoG-771310-PI2FA and by the Spanish Ministry of Economy, Industry and Competitiveness through the grant PGC2018-095832-B-I00 is acknowledged. V. Liakh, M. Luna, and E. Khomenko thankfully acknowledge the technical expertise and assistance provided by the Spanish Supercomputing Network (Red Espa\~{n}ola de Supercomputac\'{\i}on), as well as the computer resources used: the LaPalma Supercomputer, located at the Instituto de Astrof\'{\i}sica de Canarias.
\end{acknowledgements} 
 
%
   \bibliographystyle{aa} 
   \bibliography{bibtex.bib} 
%
\end{document}